\documentclass[11pt]{article}
\usepackage{amsmath}
\usepackage{amsthm}
\usepackage{amssymb}

\def\del{\partial}

\def\calM{\mathcal{M}}

\def\Rp{{R_{(+)}}}
\def\Rm{{R_{(-)}}}
\def\Pp{{P_{(+)}}}
\def\Pm{{P_{(-)}}}
\def\Qp{{Q_{(+)}}}
\def\Qm{{Q_{(-)}}}
\def\Mp{{M_{(+)}}}
\def\Mm{{M_{(-)}}}
\def\Sp{{S_{(+)}}}

\def\Tp{{T_{(+)}}}

\def\Lp{{L_{(+)}}}
\def\Lm{{L_{(-)}}}

\def\Np{{N_{(+)}}}

\def\Npm{{N_{(\pm)}}}

\def\Ip{{I_{(+)}}}
\def\Im{{I_{(-)}}}
\def\Ipm{{I_{(\pm)}}}
\def\Imp{{I_{(\mp)}}}

\def\Rpm{{R_{(\pm)}}}
\def\Ppm{{P_{(\pm)}}}
\def\Qpm{{Q_{(\pm)}}}

\def\Pmp{{P_{(\mp)}}}
\def\Qmp{{Q_{(\mp)}}}

\def\IPp{{I_{(P_+)}}}
\def\IPm{{I_{(P_-)}}}
\def\IPpm{{I_{(P_{\pm})}}}
\def\IPmp{{I_{(P_{\mp})}}}
\def\IQp{{I_{(Q_+)}}}
\def\IQm{{I_{(Q_-)}}}
\def\IQpm{{I_{(Q_{\pm})}}}
\def\IQmp{{I_{(Q_{\mp})}}}

\def\gp{{g_{(+)}}}
\def\gm{{g_{(-)}}}
\def\gpm{{g_{(\pm)}}}

\def\sig{{\sigma}}
 \def\dem{\partial_{--}}
 \def\dep{\partial_{++}}
  \def\sdem{\partial_{-}}
 \def\sdep{\partial_{+}}
 
\def\hpp{h^{++}}
\def\hmm{h^{--}}

\def\Gammap{{\Gamma^{(+)}}}
\def\Gammam{{\Gamma^{(-)}}}

\def\R{\mathcal{R}}

\newcommand{\delr}{\raise.3ex\hbox{$\stackrel{\leftarrow}{\partial }$}}
\newcommand{\dell}{\raise.3ex\hbox{$\stackrel{\rightarrow}{\partial}$}}

\newcommand{\dr}{\raise.3ex\hbox{$\stackrel{\leftarrow}{\delta }$}}
\newcommand{\dl}{\raise.3ex\hbox{$\stackrel{\rightarrow}{\delta}$}}

\def\BRST{\mathrm{BRST}}
\def\BRSTP{\mathrm{BRST}_{P}}
\def\BRSTQ{\mathrm{BRST}_{Q}}
\def\BRSTPp{\mathrm{BRST}_{P_+}}

\def\BV{\mathrm{BV}}

\def\ext{\mathrm{ext}}

\def\min{\mathrm{min}}
\def\gh{\mathrm{gh}}

\def\cM{{\cal{M}}}
\def\cN{{\cal{N}}}

\def\cG{{\cal{G}}}
\def\cO{{\cal{O}}}

\def\cG{{\cal{G}}}
\def\cR{\cal{R}}
\def\cH{\cal{H}}

\def\vphan{\vphantom{\frac{1}{2}} }

\newcommand{\auth}
{\large Vid Stojevic }
\newcommand{\kings}
{\it\small Department of Mathematics, King's College, London, UK \\  {\tt vid.stojevic@kcl.ac.uk, vid33@mac.com}}

\begin{document}




\hfill{arXiv: 0906.2028}


\vspace{20pt}

\begin{center}
{\Large{\bf Two-Dimensional Supersymmetric Sigma Models on Almost-Product Manifolds and Non-Geometry}}
\vspace{30pt}

\auth

\vspace{15pt}

\kings


\vspace{60pt}

{\bf Abstract}

\end{center}

In this paper we study the relation between the target space geometry of $(1,1)$ sigma models and the factorisation of the associated superconformal theory, at the classical level. It turns out that the superconformal currents $T_{\pm}$ factorise as $T_{\pm} = T_{\Ppm} + T_{\Qpm}$, with $T_{\Ppm}$ and $T_{\Qpm}$ conserved separately, provided that the target space admits projectors $\Ppm$ and $\Qpm$ (that is, a pair of almost-product structures),  compatible with the Riemannian structure of the target space, and covariantly constant with respect to the two torsionful connections $\nabla^{(\pm)}$ that arise naturally from the sigma model. It is a surprising result that the integrability of the projectors is not an obstruction for the associated symmetries $\delta_{\Ppm}$ and $\delta_{\Qpm}$ to form copies of the superconformal algebra. While one expects to be able to define a superconformal theory associated with a particular Riemannian submanifold defined by an integrable projector, a consequence of the above result is that there are no obstructions to defining a superconformal theory associated with non-integrable projectors. We show that this notion of non-geometry encompasses the locally non-geometric examples that arise in the T-duality inspired doubled formulations. In addition, we derive the general conditions for $(2,2)$ supersymmetry to be realised in the projective sense. This extends the relation between $(2,2)$ sigma models and bi-Hermitian geometry to the non-geometric setting. For the bosonic subsector we propose a BRST type approach to defining the physical degrees of freedom in the non-geometric scenario.


\setcounter{tocdepth}{2}
\pagebreak \tableofcontents \setcounter{page}{1}

\section{Introduction}

In this paper we study new symmetries that arise in $(1,1)$ supersymmetric sigma models on a class of almost-product target spaces characterised by $\nabla^{(\pm)}$-invariant projectors $\Ppm$, and the orthogonal complements $\Qpm$,  where $\nabla^{(\pm)}$ are connections with torsion proportional to $\pm H$, $H$ being the NS-NS three form.

It has been known since the late seventies \cite{Zumino:1979et}  that when the target manifold is K\"{a}hler, in addition to the superconformal symmetries, which are present for any Riemannian manifold, the $(1,1)$ sigma model has additional symmetries associated with the complex structure. The target space of the most general model with $(2,2)$ supersymmetry is bi-Hermitian, and was identified by Gates, Hull and Rocek in \cite{Gates:1984nk}. This geometry is characterised by the presence of two complex structures $\Ip$ and $\Im$, that obey $\nabla^{(\pm)} \Ipm = 0$, and reduces to the K\"{a}hler case when the torsion is set to zero. With the much more recent development of generalised complex geometry \cite{Gualtieri:2003dx}, it has been found that bi-Hermitian manifolds can be identified with $H$-twisted generalised K\"{a}hler manifolds  \cite{Lindstrom:2007xv}.

From the perspective of sigma models expressed in a manifestly $(1,1)$ supersymmetric manner, which we work with in this paper, it is the covariant constancy of $\Ipm$ that is responsible for the additional symmetries and the extension of the $(1,1)$ algebra to the $(2,2)$ algebra. The relation extends to general covariantly constant forms \cite{Odake:1988bh, Delius:1989fy, Howe:1991ic}: there is a symmetry associated with every $n$-form $\Lp$ obeying $\nabla^{(+)} \Lp=0$, and similarly in the anti-holomorphic $(-)$ sector. We refer to these as L-type symmetries. For $n>2$ these symmetries are non-linear, and extra care must be taken when analysing them at the quantum level  \cite{Howe:2006si}.  In general, when torsion is non-zero it is not necessary that for every form $\Lp$ there exists a corresponding $\Lm$ obeying $\nabla^{(-)} \Lm = 0$.\footnote{When $H=0$, asymmetry between the $(+)$ and $(-)$ sectors is not possible, since in this limit $\Lp = \Lm$.} For example, it is perfectly possible to have a manifold admitting only an almost-complex structure $\Ip$ obeying $\nabla^{(+)} \Ip = 0$, but no analogous $\Im$. The symmetry algebra of a sigma model on this manifold is then only enlarged to $(2,1)$ \cite{Hull:1997kk}. A different deviation from the $(2,2)$ algebra occurs when $\Ipm$ are only almost-complex, but not actually complex. In such a case the Nijenhuis tensor with one index lowered is a three-form constructed from the purely holomorphic and anti-holomorphic components of $H$, and the $(2,2)$ algebra is deformed by an L-type symmetry associated with it \cite{Delius:1989fy, Stojevic:2008qy}. The generalisation of this phenomenon to arbitrary L-type symmetries is the topic of \cite{Howe:2010az}.

While it is well appreciated that one can associates symmetries of the sigma model with $\nabla^{(\pm)}$-invariant forms in the target space, much of this paper is concerned with demonstrating that there is an analogous correspondence between symmetries of the $(1,1)$ sigma model for a target space admitting  a pair of $\nabla^{(\pm)}$-invariant projectors $\Ppm$ and their orthogonal complements $\Qpm$ obeying 
\begin{equation}
\label{eq:cov_constancy_of_PQ}
\nabla^{(\pm)} \Ppm = \nabla^{(\pm)} \Qpm = 0 \ .
\end{equation}
There is a sigma model symmetry associated with each of the four projectors, which can be expressed in terms of two covariantly constant objects $\Rp$ and $\Rm$ that square to $1$, that is, a pair of almost product structures (a review of almost-product manifolds is given in Section \ref{sec_almost_product}). Their integrability is equivalent to the vanishing of the mixed components of $H$, with respect to $P$ and $Q$. Each symmetry associated with a projector consists of two transformations; the first is obtained by acting with the projector on the standard superconformal transformations, and the second involves a non-linear transformation build from a rank-three tensor $M$ constructed from the mixed parts of $H$, which therefore vanishes for integrable projectors. Unlike the tensor $L$ used to construct the $L$-type symmetries mentioned above, the tensor $M$ is not totally antisymmetric when the indices are all lowered, and is not necessarily covariantly constant. Seeking an analogy with the symmetries associated with covariantly constant almost-complex structures described above may lead one to expect symmetries associated with non-integrable projectors to generate a deformed version of the superconformal algebra.  Surprisingly this is not the case - the $\Ppm$ and $\Qpm$ symmetries all close as superconformal algebras. The only sign of non-integrability is that for non-integrable projectors closure is only up to equation of motion terms in commutators between the holomorphic $(+)$ and antiholomorphic $(-)$ sectors.  As consistency requires, the superconformal currents $T_{\pm}$ factorise as $T_{\pm} = T_{\Ppm} + T_{\Qpm}$, with $T_{\Ppm}$ and $T_{\Qpm}$ conserved separately.


The fact that the superconformal theory factorises even when the projectors are not integrable is significant. It allows us to associate a sigma model realisation of a superconformal theory with a rank $d$ projector on a manifold $\cM$ of dimension $D > d$, rather than simply with a Riemannian target space.  An integrable projector defines a $d$-dimensional submanifold $\cN$ of $\cM$ - clearly one can  realise a superconformal theory in terms of a standard sigma model on $\cN$. Non-integrable projectors do not define submanifolds, and therefore the standard sigma model description is lost.  We will use term "non-geometric" to denote, specifically, this lack of an explicit relation between a manifold and a sigma model realisation of a conformal field theory.  This terminology is warranted on one hand, since non-integrable projectors do not define submanifolds, but could also be seen as somewhat misleading, since we still have the more elaborate description in terms of $\calM$ which \emph{is} purely geometric. A further argument in favour of adapting it is that, as we show below, our approach encompasses the "locally non-geometric" examples which arise in the T-duality covariant doubled formulations of string theory.


Next we study the conditions for $(2,2)$-supersymmetry, with both the superconformal theory and additional supersymmetries realised in the projective sense. $(2,2)$ supersymmetry realised on a compact six-dimensional Calabi-Yau corresponds to $N=2$ spacetime supersymmetry in compactification of type II string theory to four dimensions. It is clearly desirable to investigate scenarios where $N=2$ spacetime susy is realised from a non-geometric compactification. 

These conditions generalise the relation between $(2,2)$ sigma-models and bi-Hermitian geometry to the non-geometric scenario, and are briefly summarised as follows. For simplicity, we only state the conditions for the $(2,2)$ extension of the superconformal algebra associated with $\Pp$ and $\Pm$; the conditions for the analogous extensions involving the other possible combinations of projectors follow straightforwardly, and are detailed later in the paper. The $(+)$ sector side requires the target space to admit an almost-complex structure $\Ip$ which is covariantly constant with respect $\nabla^{(+)}$ and which commutes with $\Pp$.  The second supersymmetry is then associated with the covariantly constant tensor,
\begin{equation}
\IPp^i_{\  m} : = \Pp^i_j \Ip^j_{ \ k} \Pp^k_m  =  \Pp^i_j \Ip^j_{ \ m} \ ,
\end{equation}
which obeys $\IPp^2 = - \Pp$. The components of $H$ mixed with respect to $\Pp$ and $\Qp$, and with purely holomorphic or purely anti-holomorphic (w.r.t. $\Ip$) $\Pp$ and $\Qp$ indices, are all covariantly constant and deform the standard $(2,2)$ algebra by $L$-type symmetries.\footnote{In order to split the components in this way one needs to introduce an appropriate frame bundle. The details of this are in the body of the paper.} The same is true for the components of $H$ with three purely (anti)-holomorphic $\Pp$ indices. Analogous conditions come from the $(-)$ sector. Therefore, to realise the $(2,2)$ algebra, all of these components of $H$ need to vanish. The algebra closes off-shell only when $\Pp$ and $\Pm$ commute (which implies their integrability), and when $\Ip$ commutes with $\Im$; otherwise closure is only up to equation of motion terms. These conditions generalise the deformations due to the Nijenhuis tensor in the purely geometric sigma model. We note however that here the target space need not be complex in general, since there are no restrictions on the components of $H$ which are purely holomorphic or anti-holomorphic, and which lie completely in the $Q$ subspace. We also have a variety of new and interesting ways of deforming the $(2,2)$ algebra, and breaking target space supersymmetry.

Locally non-geometric backgrounds have been argued to exist due to various stringy symmetries, most notably T-duality and mirror-symmetry on manifolds with flux \cite{Shelton:2005cf, Grana:2006hr}. A T-duality invariant description of string theory can be obtained by doubling the dimension of the compact geometry of the string theory background, and restricting to the physical space via a projector whose rank is half the dimension of the doubled space  \cite{Duff:1989tf, Hull:2004in, Hull:2005hk, Dabholkar:2005ve,Hull:2006va}. The T-duality symmetry $O(d,d; \mathbb{Z} )$ is manifest in this description, and the different physical realisations related by T-duality are obtained by transforming the projectors with $O(d,d; \mathbb{Z} )$ symmetries. 

All the examples studied from this perspective are based on a doubled coset manifold \cite{Hull:2009sg}: 
\begin{equation}
\cG / \Gamma  \ ,
\end{equation}
where $\cG$ is a group of dimension $2d$, and $\Gamma$ is a discrete co-compact subgroup. Locally the space looks like the group manifold $\cG$. The physical subspace is locally of the form: 
\begin{equation}
\label{eq:doubled_group}
\cG / \tilde{\cG_L} \ ,
\end{equation}
where $\tilde{\cG_L}$ is the subgroup generated by:
\begin{equation}
\label{eq:geom_conditions1}
\tilde{X}^m   = \Pi^m_{M} L^{MN} \tilde{T}_N \ .
\end{equation}
Here $\Pi$ are projectors of rank $d$, and $L_{MN}$ is an $O(d,d)$ signature metric which only has mixed components, with respect to $\Pi$ and the conjugate projector $\tilde{\Pi}$; the coordinates $m,n$ run from $1$ to $d$, while $M,N$ run from $1$ to $2d$. Now, the condition for $\tilde{X}^m$ to form a subgroup is that the structure constants of $\cG$, $t^{mn}_{ \ \ p}$, satisfy:
\begin{equation}
\label{eq:struc_const_condition1}
t_{MN}^{ \ \ \ P} \Pi^p_{ \ P} \Pi^{mM} \Pi^{nN}  = 0 \ .
\end{equation}
In the WZW sigma-model the $H$-flux is proportional to the structure constants with the upper index lowered using $L_{MN}$. It follows from this, and the fact that $L_{MN}$ is mixed with respect to  $\Pi$ and  $\tilde{\Pi}$, that the part of $H$ with two indices along $\Pi$ and one along $\tilde{\Pi}$ vanishes.  An analogous situation occurs if we demand that
\begin{equation}
\label{eq:geom_conditions2}
\tilde{Z}_m   = \tilde{\Pi}_{mM} L^{MN} \tilde{T}_N \ .
\end{equation}
generate a subgroup, which occurs when the parts of $H$ with two indices along $\tilde{\Pi}$ and one index along $\Pi$ vanish, i.e.
\begin{equation}
\label{eq:struc_const_condition2}
t_{MN}^{ \ \ \ P} \tilde{\Pi}^p_{ \ P} \tilde{\Pi}^{mM} \tilde{\Pi}^{nN}  = 0 \ .
\end{equation}
 When both (\ref{eq:struc_const_condition1}) and (\ref{eq:struc_const_condition2}) are satisfied, $\cG$ is referred to as a Drinfel'd double and has been studied extensively in the mathematical literature \cite{Drinfeld:1986in, Klimcik:1995dy,Klimcik:1995ux}. If these conditions on the $H$-flux are not met, then the projectors are not integrable. The doubled space $\cG / \Gamma$ does not split into a product structure even locally. Our approach is clearly consistent with this, since, identifying $\Pi$ with $P$ and $\tilde{\Pi}$ with $Q$, and re-expressing the conditions for the integrability of the projectors using their covariant constancy (\ref{eq:cov_constancy_of_PQ}) results in precisely (\ref{eq:struc_const_condition1}) and (\ref{eq:struc_const_condition2}).\footnote{Supergravity theories with so-called R-fluxes \cite{Shelton:2005cf, Shelton:2006fd, Grana:2006hr} (not to be confused with Ramond-Ramond (RR) fluxes!) are argued to be effective theories of compactifications on such locally non-geometric manifolds, with the R-fluxes originating precisely from the mixed components of $H$ \cite{Hull:2009sg}.}

As group manifolds are parallelisable, compactifications on group manifolds preserve a large amount of supersymmetry. The approach to non-geometricity in this paper is from a diametrically opposite point of view, where we assume that the target space has no isometries, and that generically only $N=1$ supersymmetry is realised in the target space.  The sigma models on $\cG / \Gamma$ in the doubled approach are expressed in a manner that is manifestly covariantly with respect to the large amount of symmetry of the doubled group manifolds.  For an abelian group we have the T-duality symmetry, $O(d,d; \mathbb{Z})$, however the approach can also accommodate non-abelian dualities \cite{Giveon:1993ai, Klimcik:1995dy, Klimcik:1995ux} (these are not necessarily symmetries of string theory, but do relate different admissible backgrounds). In addition to the almost-product structure $\cR$, which defines the projectors $\Pi$ and $\tilde{\Pi}$, a second almost product structure, $S$,  is used to implement the physical constraints in the sigma mode. In the T-duality invariant formalism, both the doubled sigma model action, and the definition of $S$ involve a positive-definite $O(d,d)$ covariant metric $\cH$.  Because we do not assume the high amount of symmetry, neither $L$, nor an analogue of $\cH$ are naturally present in our sigma model. 

In the last part of the paper we outline a BRST definition of the physical degrees of freedom of a string theory based on a non-geometric CFT. For simplicity we will analyse the situation only in the bosonic subsector. In order to do this in general we need to understand two points: how to generalise the coupling of the worldsheet metric to the non-geometric situation, and how to express that only the degrees of freedom associated with two of the four projectors (say $\Pp$ and $\Pm$) are physical, while those associated with the complements ($\Qp$ and $\Qm$) are unphysical.  These issues are addressed by first gauging the symmetries associated with all the projectors, thereby introducing \emph{four} ghost fields $\{ c^{\Pp}, c^{\Pm}, c^{\Qp}, c^{\Qm} \}$ rather than the standard two. In the integrable case, and when $\Pp  = \Pm$, the gauged action can be straightforwardly expressed in closed form, simply by factorising the standard bosonic action using the projectors. For non-integrable projectors we are in general forced to expand the action order by order in the gauge fields. In the standard geometric situation the gauge fields simply parameterise the worldsheet metric, but when $\Pp  \neq \Pm$ this is no longer the case. We study the simplest non-trivial scenario, taking $\nabla^{(+)}$ and $\nabla^{(-)}$ invariant projectors to be inequivalent, but integrable and commuting.  Then it is still possible to write down a closed expression for the gauged action, but it can no longer be interpreted in terms of a coupling to a worldsheet metric. Much more serious obstacles are present when the $(+)$ and $(-)$ projectors do not commute. In this case one needs to perform an expansion not just in the gauge fields, but also in the projectors. After gauge-fixing we obtain four BRST operators associated with the four ghost fields mentioned above; concentrating on a single sector lets call these $\BRSTP$ and $\BRSTQ$.  Physical insertions are then defined as those that are $\BRSTP$ non-trivial, but are $\BRSTQ$ exact. The only allowed $\BRSTQ$ non-trivial  insertion in the path integral is $e^{iS}$ itself. The term $e^{iS}$ is both $\BRSTQ$ and $\BRSTP$ non-trivial already for an integrable almost-product structure, but in this case the partition function factorizes as $Z = Z_Q \times Z_P$ due to the integrability, and $Z_Q$ can be absorbed in the normalisation. Because no other $\BRSTQ$ excitations are allowed, we eliminate the problem of having infinitely many ground states, which would arise, for example, if we only required $\BRSTP$ invariance of physical states. For locally non-geometric backgrounds we can not disentangle the $Q$ directions and they play a non-trivial role in the definition of the theory.

A technical issue that arises here is that the BRST operator is only nilpotent when restricted to a single sector,  either $(+)$ or $(-)$. This is no longer true when both sectors are included, and gauge-fixing must in fact be handled using the more general Batalin-Vilkovisky (BV) formalism \cite{Batalin:1981jr, Batalin:1984jr} (also referred to as the antifield formalism). A familiarity with the BV formalism is therefore necessary in order to follow all the details in this part of the paper. We give a brief review in Appendix \ref{app:BV}, but the reader not familiar with the formalism should consult a more extensive reference, such as \cite{Gomis:1994he}. 




Finally, let us emphasise that throughout the paper only local effects of non-integrability are considered. First of all, the integrability of the projectors does not imply that the manifold $\calM$ is globally a product manifold. Moreover, topological non-geometric effects can also arise. These are manifested in the T-duality covariant doubled formalism by the effects of modding out by $\Gamma$ (\ref{eq:doubled_group}). Namely, the projectors even if integrable need not define sub-manifolds, having non-tensorial transformation properties (for example on overlaps between patches they could be related by transformations in a non-geometric subgroup of $O(d,d)$). While we do not discuss such "globally" non-geometric behaviour in this paper, some potentially relevant issues are encountered at the end of Section \ref{sec:non_geometric_string}

We also stress that only  classical sigma models are considered in the paper. Quantum mechanically the target space is required to be Ricci flat to the lowest order in $
\alpha'$, with further corrections as one goes up in $
\alpha'$. In the almost-product scenario, and in the bosonic subsector, the $\alpha'$ expansion has been studied in the wonderful but little known paper \cite{deBoer:1996eg}. We hope to address the supersymmetric generalisation in future work.  While there are no obvious reasons why the methods of \cite{deBoer:1996eg} should not translate straightforwardly to the supersymmetric case,  the implementation looks to be  technically very demanding due to the non-linear terms present in the supersymmetric algebra when the almost-product structures are not integrable. 


The paper is organised as follows. In Section \ref{sec:sigma_models} we review $(1,1)$ sigma models and give our conventions. Section \ref{sec_almost_product} provides a quick review of almost-product manifolds. In section \ref{sec_sym_on_a_prod_manifolds} we describe the new symmetries that occur for almost-product target spaces admitting $\nabla^{(\pm)}$-invariant projectors, and their algebras (the lengthy expression for the equation of motion terms that occur in the commutators between the $(+)$ and $(-)$  symmetry transformations are given in Appendix \ref{app:eom_terms}). In Section \ref{sec:22susy} we derive the conditions for realising $(2,2)$-supersymmetry in the projective sense.   Section \ref{sec:non_geometric_string} is concerned with generalising the coupling of the worldsheet metric to the non-geometric cases and defining the physical degrees of freedom of the non-geometric string theory. The analysis is restricted to the bosonic subsector. 
This section uses some ideas from the BRST/BV quantisation procedure, which are briefly reviewed in Appendix \ref{app:BV}. 
We give some concluding remarks in Section \ref{sec:outlook}.

\section{$(1,1)$ Sigma models}
\label{sec:sigma_models}

The action of the $(1,1)$ sigma model is given by
\begin{equation}
\label{eq:11action}
S = \int d^2 z (g_{ij} + b_{ij}) D_+ X^i D_- X^j  \ , 
\end{equation}
where $z$ are supersymmetric coordinates on the worldsheet,
\begin{equation}
z := \{ \sig^{++}, \sig^{--}, \theta^+ , \theta^- \} \ ,
\end{equation}
with $\sig$ parametrizing the bosonic and $\theta$ the fermionic directions. $X(z)$ are maps into a target space manifold $\calM$, which can be expanded as:
\begin{equation}
X^i(\sig, \theta) = \phi^i(\sig) + \theta^+ \psi_+^i(\sig) + \theta^- \psi_-^i(\sig) + \theta^+ \theta^- F^i(\sig) \ .
\end{equation}
The supercovariant derivatives obey  $ D_- D_- = i \partial_{--} $, $ D_+ D_+ = i \partial_{++} $, $\{ D_+, D_- \} = 0$, and are explicitly given by:
\begin{equation}
D_+ \equiv   \frac{\vec{\partial}}{\partial \theta^+} + i\theta^+ \partial_{++} \ \ \ , \ \ \ D_- \equiv  \frac{\vec{\partial}}{\partial \theta^-}   + i \theta^- \partial_{--}  \ .
\end{equation}
We use the double pluses and minuses to denote holomorphic worldsheet quantities, transforming in the double cover of worldsheet spinor representation, denoted by a single $+$ or $-$. The component action, after eliminating the auxiliary $F^i$ fields by their equations of motions, reads:
\begin{align}
S  = &  \int d^2 \sig \left(  \vphantom{\frac{1}{2}} (G_{ij}+b_{ij}) \dep \phi^{i} \dem \phi^{j} - iG_{ij}\nabla_{-}^{(+)}(\psi^{i}_{+}) \psi^{j}_{+} 
- iG_{ij}\nabla_{+}^{(-)} (\psi^{i}_{-}) \psi^{j}_{-}  \right. \nonumber \\
& \left. + \frac{1}{2} R^{(-)}_{ijkl} \psi^i_- \psi^j_- \psi^k_+ \psi^l_+ \right) \ .
\end{align}

The definitions and conventions are as follows: 
\begin{equation}
\Gamma^{(\pm )^m}_{ \ \  i j }  : = \Gamma^m_{  \  i j}  \pm \frac{3}{2} H^m_{ \ ij} \ ,
\end{equation}
with $H_{ijk} := db_{ijk} = b_{[ij,k]}$,\footnote{The anti-symmetrisation  convention is $[ab] := \frac{1}{2} \left( ab - ba\right)$.}
\begin{equation}
\nabla ^{(\pm)}_j v_i :=  v_{i,j} - \Gamma^{(\pm)^f}_{ \ \ j i} v_f \ ,
\end{equation}
and the Riemann tensor associated with the $\nabla ^{(\pm)}$ connections reads,
 \begin{align}
\label{eq:curvature_def}
R^{i (\pm)}_{ \ jkl} = -  2 \Gamma^{i (\pm)}_{ \ [k |j| ,l]} 
-  2 \Gamma^{m (\pm)}_{\ [k | j |} \Gamma^{i(\pm)}_{\ l ] m}  \ ,
\end{align}
 and obeys the identities:
\begin{equation}
\label{eq:R_iden1}
R^{(+)}_{ p jkm  } = R^{(-)}_{ km pj   } \ , 
\end{equation}
\begin{equation}
\label{eq:R_iden2}
R^{(+)}_{ p [ jkm ] } = -\frac{1}{2} \nabla^{(-)}_p H_{jkm} \ ,
\end{equation}
\begin{equation}
\label{eq:R_iden3}
R^{(-)}_{ p [ jkm ] } = \frac{1}{2} \nabla^{(+)}_p H_{jkm} \ .
\end{equation}
The first two indices of $R^{i (\pm)}_{ \ jkl}$  are the Lie algebra indices.

For any Riemannian target space, action (\ref{eq:11action}) is invariant under the superconformal symmetry:
\begin{equation}
\label{eq:sc_symmetry}
\delta_{\gp} X^i = a^{\gp } \dep X^i  + \frac{i}{2} D_+ (a^{\gp}) D_+ X^i \ ,
\end{equation}
where the transformation parameter $a^{\gp}$ obeys $D_- a^{\gp} = 0$. The analogous transformation in the anti-holomorphic $(-)$ sector is obtained by taking $+ \leftrightarrow -$. The $g$ in the superscript labels the symmetry, and signifies the fact that it is  the presence of the metric in the target space that is responsible for the superconformal symmetry; the $(+)$ only denotes that we are in the holomorphic sector,   and is not related to the  transformation properties of the parameter, which are easily determined from the knowledge that $X^i$ transform as a worldsheet scalars. 

In this paper we will be using ghostly transformation parameters, so that for example $a^{\gp}$ is taken to be fermionic. The reason for this is that we wish to perform the commutator calculations in preparation for BRST/BV quantisation in Section \ref{sec:non_geometric_string} (see Appendix \ref{app:BV}, where transformation parameters of gauge symmetries are promoted to ghost fields. However, a familiarity with BRST/BV is not necessary to understand most of the results in the paper, since at the level of writing down commutators and working out algebras, this flip of parity is only a technical and not a fundamental issue. It reduces to carefully working out  the sign changes in transforming between our expressions and ones that do not use ghostly parameters.

Taking into account the parity flip, the commutator between two transformations  is given by:
\begin{equation}
[ \delta_{A}, \delta_{B} ] X^i(z) = \int dz' \left\{ \frac{ \delta [ \delta_A X^i(z) ]  }{ \delta X^j (  z')}  \delta_B X^j(z')   +  \frac{ \delta [\delta_B X^i(z)]  }{ \delta X^j (  z')} \delta_A X^j (z')  \right\} \ .
\end{equation}
It is generated in the term linear in antifields in the master equation. 

The commutator of two superconformal transformations closes to a superconformal transformation,
\begin{equation}
\label{eq:sc_conmmutator}
[\delta_{\gpm} , \delta_{\gpm} ] X^i(z) =  \left[ a^{\gpm} \partial_{\pm \pm} a^{\gpm} + \frac{i}{4} D_\pm a^{\gpm} D_\pm a^{\gpm} \right] \delta_{g^\pm} X^i (z)   \ ,
\end{equation}
while the commutator between the sectors vanishes:
\begin{equation}
[\delta_{\gp} , \delta_{\gm} ] X^i(z) = 0 \ .
\end{equation}
The notation $[ \cdots] \delta_S$ will be used throughout the paper on the r.h.s. of commutator relations, and indicates that the symmetry transformation $\delta_S$ is applied with the transformation parameter denoted in the square brackets rather than original parameters $a^S$. 

The conserved currents associated with superconformal transformations are:
\begin{equation}
\label{eq:em_tensors_def}
T_{\gpm} : = g_{i j } \del_{\pm \pm} X^i D_\pm X^j \mp \frac{i}{2} H_{ijk } D_\pm X^i D_\pm X^j D_\pm X^k \ ,
\end{equation}
and obey $D_\mp T_{\gpm} = 0$ on-shell.

\section{Almost-product manifolds}
\label{sec_almost_product}

An almost-product structure\footnote{For a more detailed review of almost-product geometry the reader may like to consult \cite{Gates:1984nk}, the Appendix in \cite{Albertsson:2001dv}, or the mathematical literature \cite{Yano1, Yano2}.} is a globally defined $(1,1)$ tensor $R$ that obeys:
\begin{equation}
\label{eq:R_def}
R^i_j R^j_k = \delta^i_k \ .
\end{equation}
The manifold then holds the projectors $P$ and $Q$:
\begin{equation}
\label{eq:projectors}
P^i_j := \frac{1}{2} \left( \delta^i_j + R^i_j \right) \ \ \ , \ \ \ Q^i_j := \frac{1}{2} \left( \delta^i_j - R^i_j \right)  \  ,
\end{equation}
obeying $P^2 = P$, $Q^2 = Q$, $PQ = 0$. We take the manifold to be Riemannian, and assume compatibility with the metric:
\begin{equation}
g_{ij} = R^k_i R^m_j g_{km} \ ,
\end{equation}
which implies that $R_{ij}$, and therefore $P_{ij}$ and $Q_{ij}$, are symmetric under the exchange of indices. 
For this reason, there is no ambiguity in our writing the indices of $R$ and the projectors directly above one another. We also have  the relation:
\begin{equation}
P_{ij}+Q_{ij} = g_{ij} \ .
\end{equation}

The integrability conditions can be expressed in terms of the Nijenhuis tensor for $R$:
\begin{equation}
N^i_{(R)  j k } := 2 \left( R^m_{ [ j}  R^i_{ k ] , m} - R^i_m R^m_{ [ k , j ]} \right) \ .
\end{equation}
$Q$ is integrable if $P^i_j N^j_{(R) km} = 0$, $P$ is integrable if $Q^i_j N^j_{ (R) km} = 0$, and clearly if both $P$ and $Q$ are integrable, $N_{(R)}$ itself has to vanish.

\section{Superconformal theories associated with non-integrable projectors}
\label{sec_sym_on_a_prod_manifolds}

In this section we show that the sigma model admits additional symmetries associated with covariantly constant projectors, and derive the algebra.  

Since there are two natural connections $\nabla^{(\pm)}$, we need to distinguish between the inequivalent projectors covariantly constant with respect these, $\Ppm$ and $\Qpm$, obeying:
\begin{equation}
\label{eq:cov_constancy_P}
\nabla^{(\pm)} \Ppm  =  \nabla^{\pm} \Qpm = 0 \ .
\end{equation}
These can be expressed in terms of $\Rpm$, with the properties:
\begin{equation}
\label{eq:cov_constancy_R}
\Rpm^i_j \Rpm^j_k = \delta^i_k \ \ \ , \ \  \ \nabla^{(\pm)}  R_{(\pm)} = 0 \ .
\end{equation}
The symmetry transformations are given by: 
\begin{equation}
\label{eq:P_symmetry}
\delta_\Ppm X^i = \Ppm^i_m \left( a^\Ppm \del_{++} X^m - \frac{i}{2} D_+ a^\Ppm D_+ X^m \right)  - \frac{3}{2} i a^\Ppm M(\Ppm)^i_{ \ m p} D_+ X^{m} D_+ X^p  \ ,
\end{equation}
and depend on (\ref{eq:cov_constancy_P}) as well as the fact that the Riemann tensors (\ref{eq:curvature_def})  for $\nabla^{(\pm)}$  are pure in their Lie algebra indices with respect to $\Ppm$ $\Qpm$. Here:
\begin{align}
\label{eq:M_def}
& M( \Pp)^i_{  \ mp} := \left( - \Qp^i_r \Pp^s_m \Pp^v_p + \Pp^i_r \Qp^s_m \Qp^v_p \right) H^r_{ \ s v} \ , \\ \nonumber
& M( \Pm)^i _{  \ mp} := \left(  \Qm^i_r \Pm^s_m \Pm^v_p - \Pm^i_r \Qm^s_m \Qm^v_p \right) H^r_{ \ sv}  \ .
\end{align}
The symmetries $\delta_\Qpm$ corresponding to the $\Qpm$ projectors are obtained by exchanging $\Ppm$ and  $\Qpm$ in the above expressions.  From (\ref{eq:projectors}) it is clear that all these can be expressed in terms of the superconformal symmetry  (\ref{eq:sc_symmetry}) and the symmetries associated with $\Rpm$:
\begin{align}
\label{eq:R_symmetry}
\delta_\Rpm X^i  =  & \Rpm^i_m \left( a^\Ppm \del_{++} X^m - \frac{i}{2} D_+ a^\Ppm D_+ X^m \right) \\ \nonumber
& \mp \frac{3}{4} \Rpm^k_r \Rpm^s_m \Rpm^v_p H^r_{ \ s v} D_+ X^m D_+ X^p  \ .
\end{align}

Let us concentrate on the holomorphic sector, and take for the moment $R \equiv R_{(+)}$. Then $\nabla^{(+)} R = 0$ implies:
\begin{equation}
N^i_{(R)  j k }  \propto\left( P^i_r Q^s_j Q^v_k + Q^i_r P^s_j P^v_k \right) H^r_{  \ s v }  \ .
\end{equation}
Unlike for the almost-complex Nijenhuis tensor, $N_{(R) ijk}$ is not totally antisymmetric.  The vanishing of $N_{(R)}$ is equivalent to the mixed parts of $H$, with respect to $P$ and $Q$, vanishing. If $P$ is integrable, only  $Q^i_r P^s_j P^v_k H^r_{  \ s v } $ needs to vanish, and similarly for $Q$.

$M$ in (\ref{eq:M_def}) is clearly closely related to $N_{R}$, and it follows that on a product manifold $M_{P/Q}$ vanishes. The conformal symmetry then splits in the expected way under the projection.  We note that  the $M$ part of the transformation (\ref{eq:P_symmetry}) is not a symmetry on its own, and can certainly not be categorised as an $L$-type symmetry transformation (which we describe in the next section (\ref{eq:L_type_sym})), as $M_{ijk}$ is neither covariantly constant nor totally antisymmetric. 

It is possible that there exists only an $R_{(+)}$ obeying (\ref{eq:cov_constancy_R}), but no $R_{(-)}$ (or vice-versa), a situation which can be compared to the $(2,1)$ models mentioned in Section \ref{sec:sigma_models}.  However, while for two covariantly constant forms, $L_{(+)}$ and $L_{(-)}$, obeying $\nabla^{(+)} L_{(+)} = 0$ and   $\nabla^{(-)} L_{(-)} = 0$,  taking $L_{(+)} = L_{(-)}$ implies that $H=0$, taking $R_{(+)} = R_{(-)}$ only implies that the mixed parts of $H$ with respect to $P$ and $Q$ vanish, in other words, that the almost-product structure is integrable.

It is a somewhat surprising fact that $\delta_P$ and $\delta_Q$ form copies of the superconformal algebra, with no deformation due to the presence of the nonlinear part in (\ref{eq:P_symmetry}).  Having worked out that $[ \delta_P, \delta_P ] \sim \delta_P$ and $[ \delta_Q, \delta_Q] \sim \delta_Q$, the fact that commutator between  $\delta_P$ and $\delta_Q$ vanishes follows from the fact that $\delta_P+ \delta_Q$ is just the conformal symmetry. 

The conserved current associated with $\delta_{\Pp}$ is given by:
\begin{align}
T_{\Pp} : = &  \Pp_{i j } \del_{++} X^i D_+ X^j \\ \nonumber
& - \frac{i}{2} \left( \Pp_i^r \Pp_j^s \Pp_k^v  + 3 \Pp_i^r \Qp_j^s \Qp_k^v \right) H_{rsv} D_+ X^i D_+ X^j D_+ X^k \ ,
\end{align}
and similarly for $\delta_{\Qp}$ ($\Pp \leftrightarrow \Qp$), and in the $(-)$ sector. As consistency requires, $T_P + T_Q$ is just the superconformal current, while $T_P - T_Q$ is the conserved current associated with  $\delta_R$  (\ref{eq:R_symmetry}).  Therefore, on an almost-product manifold we have a splitting of the energy momentum tensor (\ref{eq:em_tensors_def}), such that in the non-integrable case  $P_i^r Q_j^s Q_k^v  H_{rsv}$, which is non-zero when $Q$ is not integrable, contributes to $T_P$, while $Q_i^r P_j^s P_k^v  H_{rsv}$, which is non-zero when $P$ is not integrable, contributes to $T_Q$. One can also show directly that $D_{\mp} T_{\Ppm} = D_{\mp} T_{\Qpm} = 0$ on-shell.

The derivations of the symmetry and current conservation statements and the commutators are quite lengthy, but straightforward, and  one has to make frequent use of the fact that the Riemann tensor of the torsionful connection is pure, with respect to $P/Q$, in its Lie algebra indices, and of the identities (\ref{eq:R_iden1}), (\ref{eq:R_iden2}), and (\ref{eq:R_iden3}).   In the bosonic subsector the calculations can be done quickly, and the results stated straightforwardly.  The simplifications occur due to the $H$ dependent part of the energy momentum tensor (\ref{eq:em_tensors_def}) vanishing when fermions are set to zero.

Commutators between the transformations in the $(+)$ and $(-)$ sectors  close, up to equation of motion terms that vanish if the $(+)$ and $(-)$ projectors commute. For simplicity, we demonstrate this in the bosonic subsector. The same statement is true for commutators of the $(1,1)$ superfield transformations, but the expressions are a lot more complicated to write down (and work out), and we relegate the details to Appendix \ref{app:eom_terms}.  Let us exemplify this by considering $[ \delta_{\Pp} , \delta_{\Pm} ]$. We have,
\begin{equation}
\delta_{ \Ppm}\phi^i  = a^{\Ppm} \Ppm^i_j \partial_{\pm \pm} \phi^j \ ,
\end{equation}
as symmetries of
\begin{equation}
S = \int d^2 \sigma (g_{ij} + b_{ij}) \dep \phi^i \dem \phi^j \ ,
\end{equation}
provided that (\ref{eq:cov_constancy_P}) holds. Then 
\begin{align}
\label{eq:bos_proj_pm_commutator}
[ \delta_{\Pp} , \delta_{\Pm} ]\phi^i  =   - \frac{1}{2} a^{\Pp} a^{\Pm} [ \Pp, \Pm ]^{ij} \frac{\delta S}{\delta \phi^j}  \ ,
\end{align}
where the commutator between the projectors is explicitly:
\begin{equation}
\label{eq:proj_commutator}
 [ \Pp, \Pm ]^i_{ \ j}  := \Pp^i_m \Pm^m_j - \Pm^i_m \Pp^m_j \ .
\end{equation}
 The right-hand side of (\ref{eq:bos_proj_pm_commutator}) is a symmetry due to a symmetryic-antisymmetric contraction, and is treated in the BRST/BV formalism (see Appendix \ref{app:BV}) by introducing terms non-linear in the antifields of $\phi$ in the extended action, without a need to introduce sources for the equation of motion symmetry transformations. 
  
It is obvious from (\ref{eq:projectors}) that commutators between different projectors are related: 
\begin{align}
\label{comm_equalities}
& [ \Pp, \Pm ] = [\Pm, \Qp ] =  [ \Qm, \Pp ]  \ \ \ , \ \ \   [ \Qp, \Qm ] = [ \Pm , \Qp ] =  [ \Qm , \Pp ]  \\ \nonumber
& [ \Qp,  \Pm ] = [ \Pp, \Qm ]  \ \ \  ,  \ \ \  [ \Pp , \Pm ] = [ \Qp, \Qm ]  \ .
\end{align}

Without further information projectors in the $(+)$ and $(-)$ sectors are only related by:
\begin{equation}
\Pp + \Qp = \Pm + \Qm   \ .
\end{equation}
The simplest possibility which renders vanishing commutators is that:
\begin{equation}
\label{eq:projectors_same}
\Pp = \Pm \ \ \  , \ \ \  \Qp = \Qm \ .
\end{equation}
As explained above, this implies that all the projectors are integrable. It is also easily checked that (\ref{eq:projectors_same}) can be equivalently stated as:
\begin{equation}
\Pp^i_j \Qm^j_k = 0 \ , 
\end{equation}
which expresses the fact that $\Pp$ is pure in the $\Pm$ indices and has no $\Qm$ directions, while $\Qm$ is pure in $\Qp$ indices and has no $\Pp$ directions (and vice versa).

The vanishing of the commutators is a weaker assumption than $\Pp = \Pm$, and can be equivalently thought of as $\Pp$ having no mixed part in the $\Pm$ and $\Qm$ indices (and so on for the other combinations). However, even when $\Pp \neq \Pm$, it can still be shown that the vanishing of the commutator implies integrability for the projectors. This can be seen, for example, by rewriting:
\begin{equation}
\label{eq:d_PpPm_comm}
 [ \Pp, \Pm ]_{[ij,k]} = 0 \ ,
 \end{equation}
and the other combinations in (\ref{comm_equalities}) in terms of $H$. From the resulting equations it follows straightforwardly that $H$ must be pure in its indices with respect to all the projectors.  We also note that when $\Pp \neq \Pm$ and $[ \Pp, \Pm ] = 0$, the geometry admits a set of additional projectors, $\Ppm^i_j \Pmp^j_k$, $\Ppm^i_j \Qmp^j_k$, and $\Qpm^i_j \Qmp^j_k$. These are not covariantly constant under either $\nabla^{(\pm)}$ or the Levi-Civita connection.

In the scenario when the projectors are not integrable,  all four combinations 
\begin{equation}
\label{eq:sconf_alg_copies}
\{ \delta_{\Pp} ,  \delta_{\Pm} \} \ \ \  ,  \ \ \  \{ \delta_\Pp, \delta_{\Qm} \}  \ \ \   , \ \ \    \{ \delta_\Pm, \delta_{\Qp} \} \ \ \ , \ \ \ \{ \delta_{\Qp} ,  \delta_{\Qm} \}  \ ,
\end{equation}
 realise copies of the $(1,1)$ superconformal theory, and there is no a priori reason to assume that the ranks of the projectors in the $(+)$ and $(-)$ sectors are related. Stated another way, $\Pp$ and $\Pm$ is not a preferred pairing any more than $\Pp$ and $\Qm$, without additional input relating $\Rp$ and $\Rm$. 

\section{Conditions for $(2,2)$ supersymmetry}
\label{sec:22susy}

In addition to the superconformal symmetry, the sigma model possesses additional symmetries if there are forms present in the target space which are covariantly constant with respect to the $\nabla^{(\pm )}$ connections  \cite{Odake:1988bh, Delius:1989fy, Howe:1991ic}. For an  $n$-form $\Lp_{ i_1 i_2 \cdots i_n}$ obeying $\nabla^{(+)} \Lp  = 0$, the symmetry transformation is obtained by raising one of the indices:
\begin{equation}
\label{eq:L_type_sym}
\delta_{\Lp} X^i = a^{\Lp} \Lp^{ i}_{ \  i_2 i_3 \cdots i_n} D_+ X^{i_2} D_+ X^{i_3} \cdots D_+ X^{i_n}  \ ,
\end{equation}
The conserved currents:
\begin{equation}
T_{\Lp} :=  \Lp_{ i_1  i_2 \cdots i_n} D_{+} X^{i_1} D_{+} X^{i_2} \cdots D_{+} X^{i_n}  \ ,
\end{equation}
obey $D_{-} T_{\Lp} = 0$ on-shell. We have analogous expressions for $L$-type symmetries in the $(-)$ sector.

The simplest example is that of a symmetry related to a $\nabla^{(+)}$ invariant almost-complex structure $\Ip$:
\begin{equation}
\label{eq:alm_cpx_stru_sym}
\delta_{\Ip} X^i = a^{\Ip}  \Ip^{ k}_{  \ m} D_+ X^m \  .
\end{equation}
For a pair of such structures $\Ipm$, $\nabla^{(\pm)} \Ipm = 0$, the algebra of $\delta_{g }$ and $\delta_{\Ipm}$ is given by,
\begin{align}
\label{eq:almost_cpx_algebra}
[ \delta_{\Ipm} , \delta_{ \Ipm} ]  X^i  & =  - [ i a^{\Ipm} a^{\Ipm} ] \delta_{\gpm } X^i   + [a^{\Ipm} a^{\Ipm} ] \delta_{ \Npm} X^i \\ \nonumber
[ \delta_{\Ipm} , \delta_{ \Imp} ]  X^i &=   \frac{1}{2}  a^{\Ipm} a^{\Imp}  [ \Imp, \Ipm ]^{ik} \frac{\delta S}{\delta X^k} \\ \nonumber
[ \delta_{\gpm } , \delta_{\Ipm} ] X^i & = [ a^{\gpm} \dep a^{\Ipm} - \frac{1}{2} a^{\Ipm} \dep a^{\gpm} + \frac{i}{2} D_+ a^{\gpm} D_+ a^{\Ipm}   ] \delta_{\Ipm} X^i  \ ,
\end{align}
with the commutators not listed vanishing.  Here 
\begin{equation}
 [ \Imp, \Ipm ]^i_{ \ k}  := \Imp^i_{ \ j} \Ipm^j_{ \ k} - \Ipm^i_{ \ j} \Imp^j_{ \ k} \ , 
\end{equation}
and $\delta_{\Npm}$ are L-type symmetries (\ref{eq:L_type_sym}) associated with the Nijenhuis tensors of $\Ipm$ \cite{Delius:1989fy}:
\begin{equation}
\label{eq:nijenhuis1}
\Npm^k_{ \ ij} : =  \Ipm^k_{ \ m} \Ipm^m_{ [ i , j ] } +  \Ipm^m_{ \ [ i} \Ipm^k_{ \ j ] , m} \  . 
\end{equation}
When both of the almost-complex structures are integrable, $\Npm = 0$, and (\ref{eq:almost_cpx_algebra}) is the standard $(2,2)$ algebra, as expressed in $(1,1)$ superfields. Using $\nabla^{(\pm)} \Ipm = 0$ we obtain:
\begin{equation}
\label{eq:nijenhuis2}
\Npm_{ijk} = \mp \frac{3}{2} \left( H_{ijk} - 3 \Ipm^l_{ \ [ i } \Ipm^m_{ \ j} H_{ k ] lm }  \right) \ ,
\end{equation}
which expresses that $\Npm$ is proportional to the $(3,0) + (0,3)$ component of $H$, which respect to $\Ipm$. When $H \neq 0$ the existence of an $\Lp$ obeying  $\nabla^{(+)} \Lp  = 0$ does not imply the existence of an $\Lm$ obeying  $\nabla^{(-)} \Lm = 0$. For example, the $(2,2)$ algebra must be distinguished from the $(2,1)$ algebra which is realised when only one of the complex structures is present \cite{Hull:1997kk}.


When both the almost-product structures, $\Rpm$ (\ref{eq:R_def}), and the almost-complex structures, $\Ipm$, are present, we have in general a large set of symmetries associated with the various possible projections of $\Ipm_{ij}$.  In the rest of this section we will study the symmetry algebra generated by the symmetries associated with the following projections of $\Ipm$:
\begin{equation}
\label{eq:projected_alm_cpx}
 \IPpm^i_{ \ j}   := \Ppm^i_l \Ipm ^l_{ \ m} \Ppm^m_{ \ j}  \ \  \ , \ \ \    \IQpm^i_{ \ j}   := \Qpm^i_l \Ipm ^l_{ \ m} \Qpm^m_{ \ j}  \ ,
\end{equation}
together with $\delta_{\Ppm}$ and $\delta_{\Qpm}$. In particular, we will derive the conditions for the four subsets:  
\begin{align}
\label{eq:possible22algebras}
& \{ \delta_{\IPp} ,  \delta_{\Pp},  \delta_{\IPm},  \delta_{\Pm}  \} \ \ \  ,  \ \ \  \{ \delta_{\IQp} ,  \delta_{\Qp},  \delta_{\IQm}, \delta_{\Qm} \}  \\ \nonumber 
& \{ \delta_{\IPp} , \delta_{\Pp} , \delta_{\IQm} , \delta_{\Qm} \}  \ \ \ ,  \ \ \ \{ \delta_{\IPm} , \delta_{\Pm} , \delta_{\IQp} , \delta_{\Qp} \}  \ ,
\end{align}
to generate copies of the $(2,2)$ algebra.  Clearly, each of these is an extensions of a superconformal sub-algebra listed in (\ref{eq:sconf_alg_copies}).

The simplifying assumption in our analysis is that the mixed components of $\Ipm$ vanish.   With these components turned on we would be getting into a rather more complicated scenario, where the relations
\begin{equation}
\label{eq:IPpm_conditions}
\IPpm^2 = - \Ppm \ \ \ \mathrm{and} \ \ \  \IQpm^2 = - \Qpm 
\end{equation}
would not be valid. This would in turn imply that the commutators $[ \delta_{\IPpm}, \delta_{\IPpm} ] $ generate additional symmetries associated with the mixed components of $\Ipm$, and  would prevent any of the subsets in (\ref{eq:possible22algebras}) from realising $(2,2)$ algebras. The relations (\ref{eq:IPpm_conditions})  are equivalently expressed as:
\begin{equation}
 \Ipm^2 = - 1 \ \ \  , \ \ \  [ \Rpm, \Ipm ]  = 0  \ .
\end{equation}
While $\IPpm$ and $\IQpm$ are in general not almost-complex structures (\ref{eq:IPpm_conditions}),  the structures 
\begin{equation}
\Rpm^{i}_m \Ipm^m_ j 
\end{equation}
are.

The commutators $[ \delta_{\IPpm}, \delta_{\IPmp} ] $ vanish up equation motion terms:
\begin{equation}
\label{eq:pm_comm1}
[ \delta_{\IPpm}, \delta_{\IPmp} ]  X^i - \frac{1}{2} a^{\IPpm} a^{\IPmp} [ \IPpm, \IPmp ]^{ik} \frac{\delta S}{\delta X^k}  \ ,
\end{equation}
and we have the analogous result for $[\delta_{\IQpm}, \delta_{\IQmp}]$, $[\delta_{\IPpm}, \delta_{\IQmp}]$, and $[\delta_{\IQpm}, \delta_{\IPmp}]$.  The same is true for the commutators between $\delta_{\Ppm}$ and $\delta_{\IPmp}$, which are given by:
\begin{align}
\label{eq:pm_comm2}
[ \delta_{\Ppm}, \delta_{\IPmp} ]X^i =  &  - \frac{i}{4} D_+a^{\Ppm} a^{\IPmp}  [ \Ppm, \IPmp ]^{ik} \frac{\delta S}{\delta X^k}  \\ \nonumber
& + \frac{i}{2} a^{\Ppm} a^{\IPmp} \left\{  [ \Ppm,  \IPmp ]^{ik} D_+ \left(  \frac{\delta S}{\delta X^k} \right) 
 + \left( [ \IPmp, \Ppm ]^{ij} \Gamma^{(+) k}_{ \ \  \ \  j m} \right.   \right.  \\ \nonumber
 & \left. \left. + 3 M( \Ppm )^i_{ \ mj} \IPmp^{jk} + 3 \IPmp^i_{ \ j} M( \Ppm)^{jk}_{ \ \ m} \right) D_+ X^m \frac{\delta S}{\delta X^k}  \right\} \ .
\end{align}
Again,  $[\delta_{\Qpm}, \delta_{\IQmp}]$, $[\delta_{\Qpm}, \delta_{\IPmp}]$, $[\delta_{\Ppm}, \delta_{\IQmp}]$ are all be obtained by making the obvious replacements in the above expression. 

The important point to take away from the above results is that the symmetry commutators between the $(+)$ and $(-)$ sectors involving the almost-complex structure always vanish on-shell, and vanish off-shell provided that the appropriate structures commute, i.e. $[ \delta_{\IPp}, \delta_{\IPm} ]  = 0$ provided that $[ \IPp, \IPm ]  = 0$,  $[ \delta_{\Pp}, \delta_{\IPm} ] = 0 $ provided that $[ \IPm, \Pp ] = 0$, and so on for the other combinations. Similarly to the situation encountered in the context of equation (\ref{eq:d_PpPm_comm}), the vanishing of the last line in (\ref{eq:pm_comm2}) can be shown to follow from $[ \IPmp, \Ppm ]_{[ij,k]} = 0$. Therefore, irrespective of whether the various structures commute or not, there are no obstructions to realising the $(2,2)$ algebras on-shell from the commutators between the $(+)$ and $(-)$ sectors.

To analyse the commutator relations within the $(+)$ sector, it will be useful to introduce a frame which splits with respect to $\Ip$, and also with respect to $\Pp$ and $\Qp$. Similarly, when studying the $(-)$ sector we will wish to split the indices with respect to the $\Pm$, $\Qm$ and $\Im$. To keep the notation uncluttered, we give the detailed results for the $(+)$ sector only, as the results in the $(-)$ sector follow straightforwardly. 

We denote  the tangent space indices in the $\Pp$ direction using capital letters from the beginning of the alphabet, are further split them using the almost-complex structure $\Ip$ as:
\begin{equation}
\label{eq:split_holom1}
A, \overline{A}, B, \overline{B}, \cdots  
\end{equation}
The tangent frame indices in the $\Qp$ direction denoted using capital letters from the middle of the alphabet:
\begin{equation}
\label{eq:split_holom2}
I, \overline{I}, J, \overline{J}, \cdots 
\end{equation}

The commutator between $\delta_{\IPp}$ and $\delta_{\IPp}$  is given by:
\begin{align}
[ \delta_{\IPp}, \delta_{\IPp} ] X^i =  -i \left[ a^{\IPp} a^{\IPp} \right] \delta_{ \Pp} X^i + \left[ a^{\IPp} a^{\IPp} \right] \delta_{\Mp} X^i \ .
\end{align}
Here $\delta_{\Mp}$ is an $L$-type symmetry, where the $\nabla^{(+)}$ invariant 3-form is given by:
\begin{equation}
\Sp_{ijk} =  \Pp^m_{ i} \Pp^p_m \Pp^r_j \Np_{mpr} +   \frac{9}{2} \Qp^m_{[p} \Pp^p_{j} \Pp^r_k \left( - H_{mpr} + H_{svm} \Ip^s_{ \ p} \Ip^v_{ \ r} \right)   \ .
\end{equation}
It follows that a necessary condition for $\{ \delta_{\IPp} \delta_{\Pp} \}$ to generate the $(2,2)$ algebra is $\Sp_{ijk} = 0$. Going to the frame (\ref{eq:split_holom1}), (\ref{eq:split_holom2}), one can see that $M$ consists of the following components of $H$:
\begin{align}
\label{eq:inv_H_components1}
H_{ABC}  \ \ \  , \ \ \  H_{AB I}  \ \ \ , \ \ \  H_{ A B  \overline{I}}   \ , 
\end{align}
as well as the complex conjugates, all of which are therefore separately covariantly constant. The first two are simply components of $\Np$ (\ref{eq:nijenhuis2}), the first entirely in the $\Pp$ subspace,  and the second having two legs along $\Pp$ and one along the $\Qp$ direction.  On the other hand, $H_{ A B  \overline{I}}$ (and the complex conjugate), do not belong to $\Np$, but are nevertheless $\nabla^{(+)}$ invariant. 

The commutator between $\delta_{\IPp}$ and $\delta_{\Pp}$  is given by:
\begin{align}
[ \delta_{\IPp}, \delta_{\Pp} ] X^i = &  \left[  a^{\Pp} \dep a^{\IPp}  - \frac{1}{2} a^{\IPp} \dep a^{\Pp} + \frac{i}{2} D_+ a^{\Pp} D_+ a^{\IPp} \right] \delta_{\IPp} X^i \\ \nonumber & + \left[ a^{\IPp} a^{\Pp} \right] \delta_{\Tp} X^i \ .
\end{align}
Here the obstruction to the $(2,2)$ algebra is $\delta_{\Tp}$. This is a non-linear generalisation of $\delta_{\Pp}$, which has the form:
\begin{equation}
\delta_{\Tp} X^i = a^{\Tp} \Tp^i_{(1) j k } \dep X^j D_+ X^k + i D_+ a^{\Tp}  \Tp^i_{(2) j k } D_+ X^j D_+ X^k  \ .
\end{equation}
Analyzing this type of symmetry fully is beyond what we wish to do here, but fortunately the tensors $\Tp_{(1)}$, $\Tp_{(2)}$ vanish provided that the components (\ref{eq:inv_H_components1}) are zero. 

At this stage we have derived all the conditions for  $\{ \delta_{\IPp} \delta_{\Pp} \}$ to generate the $(+)$ sector of the $(2,2)$ algebra. The corresponding conditions for the $(-)$ sector, as well as for all the sub-algebras in (\ref{eq:possible22algebras}), can be obtained by making the obvious replacements in the above commutators. 


In conclusion, let us summarise the conditions for $\{ \delta_{\IPp},  \delta_{\Pp}, \delta_\IPm, \delta_{\Pm}  \}$ to generate a copy of the $(2,2)$ algebra:
\begin{itemize}
\item $\nabla^{(\pm)} \Rpm = \nabla^{(\pm)} \Ipm = 0 \ . $
\item $ [\Rpm, \Ipm ] = 0 \ . $
\item The components of the Nijenhuis tensor $\Np$ of $\Ip$ (or equivalently, the purely holomorphic and anti-holomorphic components of $H$) with two legs along $\Pp$ and one along $\Qp$, and with all three legs in the $\Pp$ direction, need to vanish.  The analogous statement with $(+) \rightarrow (-)$.
\item The components of $H$ with two holomorphic legs, with respect to $\Ip$, in the $\Pp$ direction and one anti-holomorphic leg in the $\Qp$ direction need to vanish, as do the complex conjugate components. The analogous statement with $(+) \rightarrow (-)$.
\end{itemize}

These conditions are not strong enough to imply that $\Ppm$ are integrable, since, for example, there are no restriction on the components $ \Qp_i^m \Pp_j^r \Pp_k^s H_{mrs}$ which are mixed holomorphic/anti-holomorphic in the two $\Pp$ directions.  Also, unlike the conditions for realising the $(2,2)$ algebra in  (\ref{eq:almost_cpx_algebra}), here we do not require $\Npm = 0$, since not all the purely (anti)-holomorphic  components of $H$ are required to vanish.

The commutator between $\delta_{\IPp}$ and $\delta_{\IQp}$ generates $L$-type symmetries constructed from the $\nabla^{(+)}$ invariant components of $H$ given in (\ref{eq:inv_H_components1}), and not surprisingly, since it is symmetric between $\Qp$ and $\Pp$, it also generates L-type symmetries constructed from the components:
\begin{align}
\label{eq:inv_H_components2}
H_{IJK}  \ \ \  , \ \ \  H_{IJ A}  \ \ \ , \ \ \  H_{ I J  \overline{A}}   \ , 
\end{align}
and their complex conjugates, so that these also need to be $\nabla^{(+)}$ covariantly constant. 

It turns out that the requirement for all of the mixed components of $H$ with respect to $\Pp$ and $\Qp$ to vanish (i.e. for the almost-product structure to be integrable), is equivalent to the integrability of both almost-complex structures $\Ip^i_{ \ j}$ and $ \Rp^{i}_m \Ip^m_ j$. Again, an analogous statement is true in the $(-)$ sector. It follows that a necessary condition for \emph{all} the combinations in  (\ref{eq:possible22algebras}) to generate copies of the $(2,2)$ algebra is that both almost-product structures $\Rp$ and $\Rm$ are integrable.

\section{The non-geometric string}
\label{sec:non_geometric_string}

In section \ref{sec_sym_on_a_prod_manifolds} we showed that the algebra of the projected conformal symmetries (\ref{eq:P_symmetry}) is just the superconformal algebra, even when the projectors are not integrable, up to equation of motion terms that vanish for integrable projectors (or equivalently when all the projectors commute). In order to understand the non-geometric conformal theories as defining a string theory it is necessary, first, to understand how to generalise the coupling to a worldsheet metric. Furthermore, we need to pick the physical directions, which in the following we take to be defined by $\Pp$ and $\Pm$, and then define a procedure that eliminates the non-physical degrees of freedom associated with $\Qp$ and $\Qm$ directions. Working with the supersymmetric model involves significant technical complications, and we restrict the discussion in this section to the bosonic subsector.   We will also drop the double $\pm$ notation since there will be no reference to worldsheet spinors.

Let us first review how to re-derive the standard bosonic string theory action,
\begin{equation}
\label{eq:bosonic_string_action}
 S = - \int d^2 \sig \left[  \sqrt{-\gamma} \gamma^{\alpha \beta } g_{i j }  \partial_{\alpha} \phi^i \partial_{\beta} \phi^j  + \epsilon^{\alpha \beta} b_{ij} \partial_{\alpha} \phi^i \partial_{\beta} \phi^j  \right]   \  , 
 \end{equation}
 by gauging the conformal symmetries, 
 \begin{equation}
\label{eq:bos_conf_transformation}
\delta_{\gpm} \phi^i = a^{\gpm} \partial_{\pm} \phi^i \ .
\end{equation}
The reason for attacking the non-geometric string via gauging is that is that any direct geometric approach seems unlikely without somehow extending the framework set out in this paper.

First let us consider the gauging in just the holomorphic $(+)$ sector, whereby the ghostly holomorphic parameter in (\ref{eq:bos_conf_transformation}) $a^{\gp}$, obeying $\sdem a^{\gp} = 0$, is promoted to a gauge theory ghost with full dependence on worldsheet coordinates: $c^{\gp}$.\footnote{The notation that we follow is that the holomorphic (ghostly) parameters are denoted by $a$, while the fully local ghosts are denoted by $c$.} Then the action,
\begin{equation}
\label{eq:hol_string_action}
S_{\mathrm{hol}} = \int d^2  \sig \left[   (g_{ij} + b_{ij}) \sdep \phi^i \sdem \phi^j - 2  \hpp T_{++} \right] \ ,
\end{equation} 
 where  we introduce the gauge field $\hpp$, and 
 \begin{equation}
  T_{++} : = g_{ij} \sdep \phi^i \sdep \phi^j \ ,
 \end{equation}
 is invariant under the gauge transformation:
\begin{align}
\label{eq:hol_BRST}
\delta_{\gp} \phi^i  = &  c^{\gp} \sdep \phi^i  \ , \\ \nonumber
\delta_{\gp} \hpp  = &  \frac{1}{4} \sdem c^{\gp}  + \sdep \hpp c^{\gp}  - \hpp \dep c^{\gp}  \ .
\end{align}

While we know the gauge symmetries of   (\ref{eq:bosonic_string_action}) from standard geometric considerations, it is not obvious how to obtain (\ref{eq:bosonic_string_action})  by gauging the conformal symmetries in both the holomorphic and anti-holomorphic sectors simultaneously. Here we simply state the result, which can be obtained elegantly from a bi-Hamiltonian formulation \cite{Hull:1993kf, Stojevic:2006pq}. Namely, the action:
\begin{align}
\label{eq:string_hol_ahol_gauged}
S_{\mathrm{string}} =&  \int d^2 \sig   \frac{1}{1- 4 \hpp \hmm} \left[  \vphantom{\frac{1}{2}} (1+4\hpp\hmm ) g_{ij} {\sdep} \phi^i \sdem \phi^j  - 2 \hpp T_{++} - 2 \hmm T_{--} \right]   \\ \nonumber
& +   \int d^2 \sig b_{ij} \sdep \phi^i \sdem \phi^j \ ,
\end{align}
is invariant under (\ref{eq:hol_BRST}) (extended to both sectors). This provides us with the particular parametrisation of the metric:
\begin{equation}
\label{eq:gf_metric}
\gamma_{\alpha \beta} = 
\Lambda( \sig) \left( \begin{array}{cc}
2 \hmm &  \frac{1}{2} ( 1 + 4 \hmm \hpp)  \\
 \frac{1}{2} ( 1 + 4 \hmm \hpp)  & 2 \hpp
\end{array} \right) \ .
\end{equation}
$\Lambda$, being the overall scale of the worldsheet metric, drops out of the action, as must happen due to Weyl invariance of  (\ref{eq:bosonic_string_action}).

For (\ref{eq:hol_string_action}) gauge fixing can be performed using the standard BRST techniques, with the BRST symmetry given by (\ref{eq:hol_BRST}) together with:
\begin{equation}
\delta_{\gp} c^{\gp} = c^{\gp} \sdep c^{\gp}  \ .
\end{equation}
For the action (\ref{eq:string_hol_ahol_gauged}) the situation is not as simple, since the gauge symmetries close only up to equations of motion, and the BRST symmetry, constructed by extending the $(+)$ sector results just given to the $(-)$ sector in the obvious way, is only nilpotent on-shell. It is therefore necessary to use the full Batalin-Vilkovisky (BV) formalism.

The minimal solution to the master equation is:
\begin{align}
\label{eq:string_gauged_extended_action}
S_{\min} =&  S_{\mathrm{string}} + \int d^2 \sig    \left[  \vphantom{\frac{1}{2}}  \frac{1}{1- 4 \hpp \hmm} \phi^*_i \left( \vphan c^{\gp} \sdep \phi^i +  c^{\gm} \sdem \phi^i \right. \right. \\ \nonumber 
& -\left. 2  \hmm c^{\gp} \sdem \phi^i - 2 \hpp c^{\gm} \sdep \phi^i \vphan \right) 
 +   h^*_{++} \left( \frac{1}{4} \sdem c^{\gp}  + \sdep \hpp c^{\gp}  - \hpp \dep c^{\gp}   \right)  
\\ \nonumber
& + h^*_{--} \left( \frac{1}{4} \sdep c^{\gm}  + \sdem \hmm c^{\gm}  - \hmm \dep c^{\gm}   \right) 
  \left. + c^*_{\gp} c^{\gp} \sdep c^{\gp} + c^*_{\gm} c^{\gm} \sdem c^{\gm}  \vphantom{\frac{1}{2}} \right] \ .
\end{align}
Gauge fixing is performed by making a flip between the $h$ fields and antifields, and then making the simple choice of Lagrangian submanifold, namely setting all the antifields to zero. One traditionally refers to the new field as $b$, so following this notation, the substitution to make in (\ref{eq:string_gauged_extended_action}) is  \cite{Stojevic:2006pq}:
\begin{equation}
\label{eq:string_gauge_fixing}
h^*_{\alpha \beta} \rightarrow b_{\alpha \beta} \ \ \ , \ \ \ h^{\alpha \beta} \rightarrow - b_*^{\alpha \beta} \ .
\end{equation}
The field-antifield flip can be thought of as a "large" canonical transformation, and is sufficient to obtain an $S_0$ with no local invariances. This type of gauge fixing would not work for a general gauge theory,  but is sufficient here essentially because the gauge fields $h$ are not propagating. Keeping the antifields as sources for the BRST transformations demonstrates immediately that the BRST transformations are only nilpotent on-shell, since the extended action is highly non-linear in $b_*$. This is not the case if one restricts to only the holomorphic sector, as is obvious by setting all the $b_{--}$ and $c^{\gm}$ fields, as well as their antifields, to zero. It is a neat result that in the extended gauge-fixed action it is the $b^*$ antifields that parameterise the worldsheet metric.

In order to generalise this approach and define the non-geometric string theory it is necessary to gauge the symmetries associated with all four projectors: $\Pp$, $\Pm$, $\Qp$ and $\Qm$. Amongst other things this implies that the number of ghost fields is doubled from two to four.  In what follows we will demonstrate the gauging of $\delta_{\Pp}$ and $\delta_{\Pm}$. $\delta_{\Qp}$ and $\delta_{\Qm}$ need to be gauged in exactly the same manner, but we will not spell this out below in order to minimise clutter in the formulas.

Gauging only a single sector is straightforward. We have:
\begin{equation}
S_{\Pp   \mathrm{hol}} = \int d^2  \sig \left[   (g_{ij} + b_{ij}) \sdep \phi^i \sdem \phi^j - 2  \hpp T_{\Pp} \right]  \ ,
\end{equation}
where
\begin{equation}
T_{\Pp}  : =   \Pp_{ij} \sdep \phi^i \sdep \phi^j \ .
\end{equation}
The gauge symmetry is given by:
\begin{equation}
\delta_{\BRSTPp } \phi^i  =  \Pp^i_j c^{\Pp} \sdep \phi^j  \ \ \ \mathrm{and} \ \ \  \delta_{\BRSTPp} \hpp  =   \frac{1}{4} \sdem c^{\Pp}  + \sdep \hpp c^{\Pp}  - \hpp \dep c^{\Pp}  \ ,
\end{equation}
and gauge fixing is performed as in the standard case (\ref{eq:string_gauge_fixing}), with the replacement $\gp \rightarrow \Pp$.

The gauging of both $\delta_\Pp$ and $\delta_\Pm$ is obtained straightforwardly when $\Pp = \Pm$, since then there is no ambiguity in how to couple the worldsheet metric to the $P$ part of $g_{ij} = P_{ij} + Q_{ij}$ in  (\ref{eq:bosonic_string_action}). However, when $\Pp \neq \Pm$, we lack an obvious geometric approach and need to resort to gauging.  In general it is difficult to see how to obtain a closed form expression for the action, and we are only able to express the gauged action in terms of an expansion in both $\hpp$ and $\hmm$ as well as $\Pp$ and $\Pm$. The latter expansion terminates at finite order if $[\Pp, \Pm ] = 0$, so we can consider the obvious non-trivial case, namely taking $\Pp \neq \Pm$ with vanishing commutator. As discussed in Section \ref{sec_sym_on_a_prod_manifolds}, the projectors are still integrable in this case. One can check that:
\begin{align}
\label{eq:non_geom_string1}
S = & \int d^2 \sig \left[ \vphan (g_{ij} + b_{ij}) \sdep \phi^i \sdem \phi^j  - 2 \hpp  \Pp_{ ij} \sdep \phi^i \sdep \phi^j - 2 \hmm \Pm_{ij} \sdem \phi^i \sdem \phi^j  \right. \\ \nonumber
& - ( 1 - 4 \hpp \hmm )^{-1} 8 \hpp \hmm \Pp_{sj} \Pm^s_m \sdep \phi^j \sdem \phi^m  \\ \nonumber
& + \phi^*_i \left( c^{\Pp} \Pp^i_j \sdep \phi^j + c^{\Pm} \Pm^i_j \sdem \phi^j \right)  \\ \nonumber
& + ( 1 - 4 \hpp \hmm )^{-1} \phi^*_i \left( - 2 \hpp c^{\Pm} \Pm^i_s \Pp^s_j \sdep \phi^j - 2 \hmm c^{\Pp} \Pp^i_s \Pm^s_j \sdem \phi^j \right. \\ \nonumber
&\left. \vphan \left. + 4 \hpp \hmm \Pp^i_s \Pm^s_j  \left( c^{\Pp} \sdep \phi^j + c^{\Pm} \sdem \phi^j \right)  \right) \right]+ O(h^*) + O(c^*)  \ ,
\end{align}
satisfies the master equation. Here $O(h^*)$ and $O(c^*)$ are the terms involving the antifields of the gauge and ghost fields, which are of the same form as in the standard case (\ref{eq:string_gauged_extended_action}), taking $c^{\gpm} \rightarrow c^{\Ppm}$. Thus, even though we are able to write the action in closed form, it can not be interpreted in terms of coupling to a worldsheet metric since the submanifolds defined by $\Pp$ and $\Pm$ are inequivalent.

The true locally non-geometric string arises when $[\Pp, \Pm ] \neq 0$, in which case the expansion in  $\Pp$ and $\Pm$ does not terminate at finite order (at least not without any additional assumptions - see below), and we are not able to write down a closed expression for the gauged action.\footnote{The problem is manifested when using the bi-Hamiltonian method of \cite{Hull:1993kf, Stojevic:2006pq} in the inability to write down the equations of motion for the momenta in closed form, so their elimination must also be done order by order.}  In addition, we have the non-linearity in $\phi^*$ to deal with. That is, 
\begin{equation}
\int d^2\sig   \frac{1}{4}  c^{\Pp} c^{\Pm} \phi^*_i \phi^*_j [  \Pp, \Pm ]^{ij}
\end{equation}
needs to be added to the extended action in order to satisfy the master equation, and coupled to the gauge fields in an appropriate manner.  If we assume that no additional target space structures are generated by the commutator, it is necessary to equate $\Pp_{j[i} \Pm^j_{k]}$ to some part of $b_{ik}$. In the gauging procedure we still obtain the second line of (\ref{eq:non_geom_string1}) to lowest order in the gauge fields, except that $\Pp_{j[i} \Pm^j_{k]} \neq 0$.  The novelty is thus that, unlike in (\ref{eq:string_hol_ahol_gauged}) and (\ref{eq:non_geom_string1}), the $b_{ij}$ field term now couples to the gauge fields. Furthermore, $b_{ij}$ is not a tensor unless $H$ is exact, but rather has an interpretation in terms of gerbes \cite{Belov:2007qj}. This then has an interesting consequence, as it implies that the projectors should also transform non-tensorially. We leave a detailed study of these possibilities for future work.

Irrespective of our inability to obtain a closed form action when gauging the conformal symmetries associated with non-integrable projectors, we are nevertheless able to gauge-fix as in (\ref{eq:string_gauge_fixing}). For the standard bosonic string, the part of the action independent of antifields after gauge-fixing is just:
\begin{equation}
S_{\BRST} = \int d^2  \sig \left[   (g_{ij} + b_{ij}) \sdep \phi^i \sdem \phi^j   -  \frac{1}{4}  b_{++}\sdem c^{\gp}  -  \frac{1}{4}  b_{--} \sdep c^{\gm}  \right] \ .
\end{equation}
This action is still invariant under the standard string theory BRST operator; in the $(+)$ sector this is:
\begin{equation}
\label{eq:BRST_sc}
\delta_{\BRST} \phi^i  =   c^{\gp} \sdep \phi^i    \ \  \ , \ \ \   \delta_{\BRST} b_{++} =  2  T_{++}    \ \ \  , \ \ \  \delta_{\BRST} c^{\gp} = c^{\gp} \sdep c^{\gp} 
\end{equation}
 However, our interpretation of the physical content in the almost-product scenario is different, since it was not the overall conformal symmetry that was gauged, but rather the projected conformal subalgebras. After gauging the four conformal subalgebras, and gauge-fixing, we have four BRST operators:
\begin{equation}
\label{eq:BRST_Pp}
\delta_{\BRSTPp} \phi^i  =   c^{\Pp} \Pp^i_j \sdep \phi^j    \ \  \ , \ \ \   \delta_{\BRSTPp} b_{\Pp ++} =  2  T_{\Pp}    \ \ \  , \ \ \  \delta_{\BRSTPp} c^{\Pp} = c^{\Pp} \sdep c^{\Pp}  \ ,
\end{equation}
and similarly for the $(-) $ sector and the $Q$ projectors. As explained above, to analyse all these operators simultaneously, we in fact need to  bring in the full BV machinery.

Schematically, in the path integral language, the partition function is:
\begin{equation}
Z = \int [d \phi  \ d b \  d c ] \exp{i S_{\BRST} } \ .
\end{equation}
For a product geometry, $Z$ splits as: 
\begin{equation}
\label{eq:int_part_function}
Z = Z_P \times Z_Q \times Z_{\mathrm{ghost} } \ .
\end{equation}
Inserting vertex operators invariant under $\BRSTP$ but exact under $\BRSTQ$ simply leaves  $Z_Q$ as an overall factor which is ultimately absorbed in the normalisation. The $\BRSTQ$ requirement is crucial, as non-trivial $\BRSTQ$ insertions are also non-trivial in the $\BRSTP$ cohomology. In particular this would imply an infinite number of ground states.  The split  (\ref{eq:int_part_function}) no longer occurs when the projectors are non-integrable, but in order to define physical observables for the general doubled path integral in the BRST language it remains natural to require them to be non-trivial in the $\BRSTP$  cohomology but $\BRSTQ$ exact.   It is nevertheless true that the term $e^{iS}$ in the path integral is both $\BRSTP$ and $\BRSTQ$ invariant. In the geometric case the partition function factorizes, as in (\ref{eq:int_part_function}), and the $\BRSTQ$ insertion is not physically relevant, but we assert that that in the non-geometric case this single non-trivial $\BRSTQ$ insertion forms a non-trivial part in the definition of the theory, and is in fact the hallmark of non-geometricity.  

This definition of a non-geometric string theory has an immediate non-trivial consequence.  Classically the ranks of $P$ and $Q$ are arbitrary, but quantum mechanically both the $P$ and the $Q$ conformal theories need to be anomaly free in order for the definition of physical degrees of freedom to be consistent. It follows that their ranks must be equal, and that the quantum theory defined using the BRST approach automatically requires a target space with all directions doubled.

\section{Outlook}
\label{sec:outlook}

In this paper we have studied symmetries that occur in $(1,1)$ sigma models on almost-product manifolds with $\nabla^{(\pm)}$-invariant projectors. We showed that the symmetries associated with projectors still close as superconformal algebras, up to equation of motion terms, irrespective of the integrability of the projectors. The non-geometric superconformal theories then correspond to non-integrable projectors on some larger manifold. We also derived the conditions for realising $(2,2)$ supersymmetric theories in the non-geometric sense.  The language we used was manifestly $(1,1)$ supersymmetric, and it would be interesting to see how this can be made manifestly $(2,2)$ supersymmetric. It is far from obvious how this should be achieved, as in general the superconformal and $(2,2)$ algebras close only on-shell in our formulation.  Finally, we described how one can define a non-geometric string by gauging these symmetries, and the ever increasing degree of complexity that occurs as the projectors are taken to be non-integrable and non-commuting. In many ways this is only a first step. Most obviously, we only did the analysis in the bosonic subsector, and did not obtain a closed expression for the expansion in the gauge fields when the projectors are non-integrable. Further to this, it would be desirable to make the relation with the T-duality covariant doubled approach fully transparent. It seems that the central issue is finding an appropriate metric on the doubled manifold under whose $\Gamma + H$ connection the almost-product structure would be covariantly constant.


The double of an arbitrary Calabi-Yau, or a bi-Hermitian geometry should presumably provide a mirror-symmetric formulation of string theory \cite{Strominger:1996it}, but at present we are able to say very little about what such a double should be.\footnote{For a possible way to address this question in the context of topological theories, see  \cite{Stojevic:2008hc}.}  Inspired by the formulation of bi-Hermitian geometry in the generalised geometry language \cite{Lindstrom:2007xv}, a possible avenue into this suggested by the work here would be to develop a generalised formalism that combines generalised K\"{a}hler geometry with generalised product geometry (the later is defined in terms of an object on $T \oplus T^*$ that squares to unity, generalising $R$ in (\ref{eq:R_def})). It is not unrealistic to hope that a manifestly mirror symmetric formulation of string theory is most naturally expressed in terms of such a generalised complex-product hybrid.

The results in  \cite{deBoer:1996eg} could provide a direct way of understanding non-geometric compactifications of string theory from the target space perspective, since the $\beta$-functions to the lowest order in $\alpha'$ are worked out for non-integrable projectors.  Ignoring the dilaton terms, these are:
\begin{equation}
\label{eq:non_geometric_eom}
\Pp^{i}_ j R^{j(+)}_{ \ k i m } = 0  \ \ \  , \ \ \   \Pm^{i}_ j R^{j (-)}_{ \ k i m } = 0   . 
\end{equation}
Similar equations are valid if we consider the conformal theory associated with the $\Qpm$ projectors. In the integrable case, and when $\Pp = \Pm$, the Riemann tensor on the target space splits, equations (\ref{eq:non_geometric_eom}) are identical, and amount to the vanishing of the Ricci tensor of the $\nabla^{(+)}$ connection on the physical submanifold $\cN$ defined by $P$. These are just the stringy equations of motion on $\cN$ up to dilaton terms. The two equations in (\ref{eq:non_geometric_eom}) are \emph{not} equivalent for non-integrable projectors, and it would be of interest to find the doubled target space action from which they can be derived. This should then enable the "compactification" procedure yielding effective theories with R-flux \cite{Shelton:2005cf, Grana:2006hr} to be performed, at least in principle. It would also be of interest to adopt the conformal field theory approach of \cite{Halpern:1990cc, Halpern:1995js, deBoer:1996eg} to the BRST formulation described in section \ref{sec:non_geometric_string}.  In fact, an alternative to studying the non-geometric CFT using the methods of \cite{deBoer:1996eg} would be via algebraic renormalisation \cite{Howe:2006si}.  Within this approach one would used the gauged extended action derived in section \ref{sec:non_geometric_string}, but instead of implementing a gauge-fixing step would treat the antifields as sources for the (super)conformal symmetries and the gauge fields as sources for the associated conserved currents.

A much more ambitious, but fundamentally interesting question, is difficult to ignore. Within the formalism presented in this paper there is no obvious  principle preventing  the inclusion of time directions in the analysis  when the projectors are not integrable. Is it, then, possible for such a theory to be  meaningful? It is difficult to imagine what such a "fuzzy" time coordinate should mean as far as the evolution of a quantum system is  concerned. It seems that either a principle exists that renders such theories inconsistent, or a "fuzzy" and inherently stringy generalisation of time evolution in quantum mechanics is mathematically sensible.

\vskip 2cm

{\large \bf Acknowledgments}\\ [2mm]

I am especially grateful to Paul Howe for observations that initiated this project, in the context of the joined work \cite{Howe:2010az}, and for reading an earlier version the manuscript.  This work was supported in part by the DFG (German Science Foundation) and a NWO VICI grant.


\appendix

\section{The equation of motion terms in $[ \delta_{(+)} , \delta_{(-)} ]$ } 
\label{app:eom_terms}

The equation of motion symmetry arising in the $[ \delta_\Pp , \delta_ \Pm ]$ commutator is given by
\begin{align}
[ \delta_\Pp , \delta_ \Pm ] X^k = &  \frac{1}{2} a^+ a^- [ \Pp , \Pm ]^k_{ \ i} D_- D_+ ( g^{ij} S_{,j}) \\ \nonumber
& + \frac{1}{2} a^+ a^-  [ \Pp , \Pm ]^k_{ \ p}  \left(   \vphantom{\frac{1}{2}}  -  \Gammap^p_{ \  i m }  D_+X^m  D_- (g^{ij} S_{,j}) + \Gammap^p_{ \ mi} D_-X^m D_+ ( g^{ij} S_{,j})  \right. \\ \nonumber
& \left. - \Gamma^{ p}_{ \ im} D_+ D_- X^m (g^{ij} S_{ , j})  \vphantom{\frac{1}{2}} \right)  \Pp^k_p  \Mm^p_{ \ im}   \\ \nonumber
&  - \frac{3}{2}   \left(   \vphantom{\frac{1}{2}} \Pp^k_p  \Mm^p_{ \ im} + \Mm^k_{ \ mp} \Pp^p_i \right) D_- X^m D_+(g^{ij} S_{ , j }) \\ \nonumber
& - \frac{3}{2}   \left(   \vphantom{\frac{1}{2}}  \Pm^k_p  \Mp^p_{ \ im}   + \Mp^k_{ \ mp} \Pm^p_i \right) D_+ X^m D_-(g^{ij} S_{ , j })  \\ \nonumber
& +  \frac{1}{2} a^+ a^- [ \Pp , \Pm ]^k_{ \ p}  \left( \vphan \Gammap^r_{ \ sm} \Gammap^s_{ \ pi} + \Gammap^r_{ \ ps} \Gammap^s_{ \ im}   + \Gammap^r_{ \ pi , m} \right. \\ \nonumber
& \left. - 2 \Gammam^r_{ \ m [ p,i ] }  -  \Gamma^r_{ \ is} \Gammap^s_{ \ pm}  \right) D_-X^p D_+ X^m g^{ij} S_{ ,j} \\ \nonumber
&  +  \frac{1}{2} a^+ a^- [ \Pp , \Pm ]^k_{ \ p}  \left( - 3(  \Pp^k_r \Mm^r_{ \ sp} +   \Mm^k_{ \ pr} \Pp^r_s ) \Gammap^s_{ \ im}   \right. \\ \nonumber
& + 3( \Pm^k_r \Mp^r_{  \ sm} + \Mp^k_{ \ mr} \Pm^r_s ) \Gammap^s_{ \ pi} \\ \nonumber 
& \left. - 9 ( \Mp^k_{ \ ms} \Mm^s_{ \ ip} - \Mm^k_{ \ ps} \Mp^s_{ \ im} )  \right)  D_- X^p D_+ X^m g^{ij} S_{, j}  \ ,
\end{align}
where
\begin{equation}
S_{ , i} := \frac{ \delta S }{ \delta X^i}   \ .
\end{equation}
The other commutators, $[ \delta_\Pp , \delta_ \Qm ]$, $[ \delta_\Qp , \delta_ \Pm ]$,  and $[ \delta_\Qp , \delta_ \Qm ]$ are obtained in the obvious way by interchanging the projectors. As explained in Section \ref{sec_sym_on_a_prod_manifolds}, the commutativity of the projectors implies their integrability. It is for this reason that the $M$ terms vanish when the projectors commute.

\section{The Batalin-Vilkovisky formalism}
\label{app:BV}

In this appendix we describe how the Batalin-Vilkovisky (BV) \cite{Batalin:1981jr, Batalin:1984jr} prescription is used to gauge fix a classical gauge invariant action $S_0$; for a thorough introduction to the subject we refer the reader to  \cite{Gomis:1994he}. We take $S_0$ to be a functional of the fields $\phi^i(x)$, and invariant under a set of gauge symmetries, which we write as:
\begin{equation}
\label{eq:gauge_transf}
\delta \phi^i = \varepsilon^A \R_A^i \ .
\end{equation}
Repeated indices here signify  \emph{both} summation and integration, a notation which will be used throughout this section. The gauge transformations are  labeled by the capital letter index, so $\varepsilon^A$ stand for a set of transformation parameters depending on $x$. The gauge invariance of the action is expressed as:
\begin{equation}
\label{eq:inv_of_action}
\frac{\dr S_0}{ \delta \phi^i} \varepsilon^A \R_A^i =0 \ . 
\end{equation}

The first step in the gauge fixing procedure is to introduce ghost fields $c^A$, which have opposite parity to the gauge transformation parameters $\varepsilon^A$. These are grouped together with $\phi^i$ in a collective field:
\begin{equation}
\label{eq:collective_field}
\Phi^\alpha = \{ \phi^i, c^A \} \ .
\end{equation}
In addition, a set of fields $\Phi^*_\alpha$ are introduced that have opposite parity to $\Phi^\alpha$. These are referred to as \emph{antifields}.  

The next ingredient is the \emph{antibracket}, which is an odd symplectic structure on the space of fields and antifields, and can locally be put in the form:
\begin{equation}
\label{eq:antibracket_def}
(A, B) :=   \frac{ \dr A} { \delta   \Phi^\alpha } \frac{\dl B}{ \delta \Phi^*_\alpha} - 
\frac{\dr A}{ \delta \Phi^*_\alpha }  \frac{ \dl B}{ \delta   \Phi^\alpha }  \  ,
\end{equation}
acting on two objects $A$ and $B$ that depend on $\Phi$ and $\Phi^*$.  

Next, we construct an action that starts as,
\begin{equation}
\label{eq:min_solution}
S_\min = S_0 + \phi^*_i c^A \R^i_A + \cdots  \ ,
\end{equation}
with the dots completed by requiring $S_\min$ to be a solution to the \emph{classical master equation}:
\begin{equation}
\label{eq:master_equation}
(S_\min, S_\min) = 0 \ .
\end{equation}
For reasons that will become clear shortly, $S_\min$ is called the minimal solution. For a gauge algebra that is not reducible and closes on-shell the solution to the master equation is given by:
\begin{equation}
\label{eq:closed_alg_BV_action}
S_{\min} = S_0 + \phi^*_i c^A \R_A^i + c^*_C N^C_{ \ AB} c^A c^B \ ,
\end{equation}
where $N^C_{ \ AB}$  are the (possibly field dependent) structure functions of the gauge algebra.  The master equation reads,
\begin{align}
\label{eq:master_equation_expanded}
\frac{1}{2} (S_\min ,S_\min ) =  & \frac{ \dr S_0 }{\delta \phi^i} c^A \R^i_A   + \phi^*_i \left[ c^A \frac{ \dr  \R^i_A}{\delta \phi^k} c^B R^k_B + (-1)^{\epsilon_{\phi_i} \epsilon_{c^F}} \R^i_F N^F_{ \ AB} c^A c^B \right] \\ \nonumber
& + c^*_D \left[2 N^D_{ \ AF} c^A N^F_{ \ GH} c^G c^H  - c^B c^C \frac{\dr N^D_{BC} }{\delta \phi^k} c^A \R^k_A \right] = 0 \ ,
\end{align}
where $\epsilon \in \mathbb{Z}_2$ is zero when the field in its subscript is bosonic and one when it is fermionic. The term independent of antifields expresses the invariance of the action, the term proportional to $\phi^*$ the closure of the algebra, while the term proportional to $c^*$ is related to the Jacobi identity.  In reference to the standard BRST procedure,  antifields in (\ref{eq:closed_alg_BV_action}) are simply sources for BRST transformations, and the master equation expresses their nilpotence.  For gauge algebras that close up to equations of motion (open algebras), terms non-linear in the antifields are needed to obtain a solution to the master equation.\footnote{A reducible gauge theory, where the generators of gauge symmetries are not all independent, is another example of a system that can be handled using the BV, but not  the BRST formalism  \cite{Gomis:1994he}.}

The gauge fixing step consists of picking a Lagrangian submanifold in the space of fields and antifields, by which we mean that the odd symplectic structure (\ref{eq:master_equation}) vanishes when restricted to it. The obvious choice is to set all the antifields $\Phi^*$ to zero, but this leaves the standard gauge invariant action, which is clearly not a good starting point for defining the quantum theory.  So we seek a deformation away from this choice, by performing a canonical transformation, i.e. a transformation $\Phi \rightarrow \Phi^{'}$, $\Phi^* \rightarrow \Phi^{*'}$,  that preserves the antibracket (\ref{eq:antibracket_def}). It can be shown that such a transformation is generated by a fermionic function $F(\Phi, \Phi^{*'})$ of fields and antifields as:
\begin{equation}
\label{eq:canonical_transf}
\Phi^{\alpha '}  = \frac{\delta F(\Phi, \Phi^{*'}) }{\delta \Phi^{*'}_\alpha} \ \ \ ,  \ \ \ \Phi^{*}_\alpha =  \frac{\delta F(\Phi, \Phi^{*'}) }{\delta \Phi^{A} }  \ .
\end{equation}
For most purposes it is sufficient to consider a less general set of transformations when $F$ is of the form
\begin{equation}
\label{eq:special_form_fermion}
F(\Phi, \Phi^{*'}) = \Phi^\alpha  \Phi^{*'}_\alpha + \Psi(\Phi, \Phi^{*'}) \ ,
\end{equation}
where $\Psi$ is referred to as the \emph{gauge fixing fermion}.  Furthermore, it is sufficient to let $\Psi$ depend only on fields, so that the canonical transformation acts only on antifields. In this case the canonical transformations are simply:
\begin{equation}
\label{eq:gauge_fixing}
\Phi^*_\alpha \rightarrow \Phi^*_\alpha + (\Psi(\Phi), \Phi^*_\alpha ) \ \ \ ,  \ \ \ \Phi^\alpha \rightarrow \Phi^\alpha  \ .
\end{equation}
This deforms the classical action by terms coming from the antifields dependent part of the extended action, but does not generate field redefinitions. Performing field redefinitions that  preserve the canonical form of the antibracket (\ref{eq:antibracket_def}) necessitates the use of (\ref{eq:canonical_transf}), and the form (\ref{eq:special_form_fermion}) is no longer sufficient.  

The minimal solution has a global $U(1)$  "ghost" symmetry, where the $U(1)$ charges, referred to as ghost numbers, are conventionally assigned as:
\begin{equation}
\gh (\phi^i) = 0 \ \ \ , \ \ \ \gh (c^A) = 1 \ \ \ ,  \ \ \ \gh (\Phi^*_\alpha) = - \gh (\Phi^\alpha) - 1 \ .
\end{equation}
In order for a canonical transformation to preserve ghost number it is necessary that $\gh(F) = \gh (\Psi) = -1$.  We need a $\Psi$ independent of antifields, but such a fermion can't be constructed from the fields in the minimal solution, since these all have positive ghost number. To cure this, auxiliary field pairs $b^A$ and $\lambda^A$ are introduced, with $\gh (b^A) = -1$ and $\gh (\lambda^A) = 0$, which can then be used to construct an appropriate $\Psi$. The extended action with the auxiliary fields (the non-minimal solution) reads:
\begin{equation}
\label{eq:me_with_aux}
S_{\ext} = S_\min + b^*_A \lambda^A \ .
\end{equation}
After performing an appropriate canonical transformation one obtains an action which, after setting the antifields to zero, has a well defined propagator.  The antifields can be kept in the path integral expression as background fields, at the linear level acting as sources for BRST transformations. We note that, at the linear level, we have simply restated the BRST procedure, where gauge fixing is performed by adding $\delta_{\BRST} \Psi$ to the classical action.

The BV procedure is justified if the quantum theory is independent of the choice of gauge fixing fermion (other than at the singular point when $\Psi  =0$). This is most succinctly stated in terms of the quantum effective action,
\begin{equation}
\label{eq:effective_action_me2}
\Gamma [ \Phi^i_{(c)}, \Phi^*_\alpha] := -i Z[J, \Phi^*]  - J_\alpha \Phi^\alpha_{(c)} \ ,
\end{equation}
where  $Z$ is the generating functional, and:
\begin{equation}
 \Phi^\alpha_{(c)}  := -i \frac{  \dr ( \ln Z ) }{\delta J_\alpha} \ .
\end{equation}
One can show that the theory is  independent of the choice of gauge fixing fermion, provided the  effective action obeys the classical master equation
$( \Gamma, \Gamma) = 0$, where, crucially, the functional derivatives are with respect to $\Phi^i_{(c)}$. This is equivalent to the more usual formulation in terms of the quantum master equation, which we won't go into here.

In defining observables, it is crucial that the solution to the master equation defines a nilpotent operator, $\delta_{\BV}$:
\begin{equation}
\label{eq:BVoperator}
\delta_{\BV} A : = (A, S_\ext ) \ .
\end{equation}
Without any restrictions on the observables the theory would be non-unitary, since the ghost fields do not obey the spin-statistics theorem.  The natural restrictions is to require observables to  obey $\delta_{\BV} \cO   = 0$, because then one can show that their expectation values are independent of the gauge choice.  However, any observable of the form $ \cO = \delta_{\BV} F$ can be considered trivial since it is automatically closed, and it follows that physically distinct observables are classified by the  cohomology of $\delta_{\BV}$. In order to respect the classical limit, a further restriction is to require observables to have ghost number zero.  $\delta_\BV \Phi^\alpha$ is independent of antifields only if the solution to the master equation is linear in the antifields (\ref{eq:closed_alg_BV_action}), in which case  it corresponds to the standard BRST transformations. It is conventional to define the BRST operator in the general case as:
\begin{equation}
\delta_{\BRST} \Phi^\alpha =  \delta_{\BV}  \Phi^\alpha |_{\Phi^*_\alpha = 0} \ ,
\end{equation}
however, this operator then only nilpotent up to equations of motion whenever $S_\ext$ has terms nonlinear in antifields.



\begin{thebibliography}{99}

\bibitem{Zumino:1979et}
B.~Zumino, ``Supersymmetry And K\"{a}hler Manifolds,'' Phys.\ Lett.\ B
{\bf 87} (1979) 203.

\bibitem{Gates:1984nk}
  S.~J.~.~Gates, C.~M.~Hull and M.~Rocek,
  ``Twisted Multiplets And New Supersymmetric Nonlinear Sigma Models,''
  Nucl.\ Phys.\  B {\bf 248} (1984) 157.

\bibitem{Gualtieri:2003dx}
 M.~Gualtieri,
 ``Generalized complex geometry,''
 arXiv:math.dg/0401221.

\bibitem{Lindstrom:2007xv}
 U.~Lindstrom, M.~Rocek, R.~von Unge and M.~Zabzine,
 ``A potential for generalized Kaehler geometry,''
 arXiv:hep-th/0703111.

\bibitem{Odake:1988bh}
S.~Odake, ``Extension Of N=2 Superconformal Algebra And Calabi-Yau
Compactification,'' Mod.\ Phys.\ Lett.\ A {\bf 4} (1989) 557.

\bibitem{Delius:1989fy}
G.~W.~Delius and P.~van Nieuwenhuizen, ``Supersymmetric Nonlinear D
= 2 Sigma Models With Nonvanishing Nijenhuis Tensor,'' ITP-SB-89-63

\bibitem{Howe:1991ic}
P.~S.~Howe and G.~Papadopoulos, ``Holonomy groups and W
symmetries,'' Commun.\ Math.\ Phys.\  {\bf 151}, 467 (1993)
[arXiv:hep-th/9202036].

\bibitem{Howe:2006si}
  P.~S.~Howe and V.~Stojevic,
  ``On the symmetries of special holonomy sigma models,''
  JHEP {\bf 0612}, 045 (2006)
  [arXiv:hep-th/0606270].

\bibitem{Hull:1997kk}
  C.~M.~Hull,
  ``Actions for (2,1) sigma models and strings,''
  Nucl.\ Phys.\  B {\bf 509} (1998) 252
  [arXiv:hep-th/9702067].

\bibitem{Stojevic:2008qy}
  V.~Stojevic,
  ``Topological A-Type Models with Flux,''
  JHEP {\bf 0805} (2008) 023
  [arXiv:0801.1160 [hep-th]].

\bibitem{Howe:2010az}
  P.~S.~Howe, G.~Papadopoulos and V.~Stojevic,
  ``Covariantly constant forms on torsionful geometries from world-sheet and
  spacetime perspectives,''
  arXiv:1004.2824 [hep-th].

\bibitem{Shelton:2005cf}
  J.~Shelton, W.~Taylor and B.~Wecht,
  ``Nongeometric Flux Compactifications,''
  JHEP {\bf 0510} (2005) 085
  [arXiv:hep-th/0508133].

\bibitem{Grana:2006hr}
  M.~Grana, J.~Louis and D.~Waldram,
  ``SU(3) x SU(3) compactification and mirror duals of magnetic fluxes,''
  JHEP {\bf 0704} (2007) 101
  [arXiv:hep-th/0612237].

\bibitem{Duff:1989tf}
  M.~J.~Duff,
  ``Duality Rotations In String Theory,''
  Nucl.\ Phys.\  B {\bf 335}, 610 (1990).

\bibitem{Hull:2004in}
  C.~M.~Hull,
  ``A geometry for non-geometric string backgrounds,''
  JHEP {\bf 0510}, 065 (2005)
  [arXiv:hep-th/0406102].

\bibitem{Hull:2005hk}
  C.~M.~Hull and R.~A.~Reid-Edwards,
  ``Flux compactifications of string theory on twisted tori,''
  arXiv:hep-th/0503114.

\bibitem{Dabholkar:2005ve}
  A.~Dabholkar and C.~Hull,
  ``Generalised T-duality and non-geometric backgrounds,''
  JHEP {\bf 0605}, 009 (2006)
  [arXiv:hep-th/0512005].

\bibitem{Hull:2006va}
  C.~M.~Hull,
  ``Doubled geometry and T-folds,''
  JHEP {\bf 0707}, 080 (2007)
  [arXiv:hep-th/0605149].

\bibitem{Hull:2009sg}
  C.~M.~Hull and R.~A.~Reid-Edwards,
  ``Non-geometric backgrounds, doubled geometry and generalised T-duality,''
  arXiv:0902.4032 [hep-th].

\bibitem{Drinfeld:1986in}
  V.~G.~Drinfeld,
  ``Quantum groups,''
  J.\ Sov.\ Math.\  {\bf 41}, 898 (1988)
  [Zap.\ Nauchn.\ Semin.\  {\bf 155}, 18 (1986)].

\bibitem{Klimcik:1995dy}
  C.~Klimcik and P.~Severa,
  ``Poisson-Lie T-duality and Loop Groups of Drinfeld Doubles,''
  Phys.\ Lett.\  B {\bf 372}, 65 (1996)
  [arXiv:hep-th/9512040].

\bibitem{Klimcik:1995ux}
  C.~Klimcik and P.~Severa,
  ``Dual Nonabelian Duality And The Drinfeld Double,''
  Phys.\ Lett.\  B {\bf 351}, 455 (1995)
  [arXiv:hep-th/9502122].

\bibitem{Giveon:1993ai}
  A.~Giveon and M.~Rocek,
  ``On nonAbelian duality,''
  Nucl.\ Phys.\  B {\bf 421}, 173 (1994)
  [arXiv:hep-th/9308154].
  
\bibitem{Batalin:1981jr}
  I.~A.~Batalin and G.~A.~Vilkovisky,
  ``Gauge Algebra And Quantization,''
  Phys.\ Lett.\  B {\bf 102} (1981) 27.

\bibitem{Batalin:1984jr}
  I.~A.~Batalin and G.~A.~Vilkovisky,
  ``Quantization Of Gauge Theories With Linearly Dependent Generators,''
  Phys.\ Rev.\  D {\bf 28} (1983) 2567
  [Erratum-ibid.\  D {\bf 30} (1984) 508].

\bibitem{Gomis:1994he}
  J.~Gomis, J.~Paris and S.~Samuel,
  ``Antibracket, antifields and gauge theory quantization,''
  Phys.\ Rept.\  {\bf 259}, 1 (1995)
  [arXiv:hep-th/9412228].

\bibitem{deBoer:1996eg}
  J.~de Boer and M.~B.~Halpern,
  ``Unified Einstein-Virasoro master equation in the general non-linear  sigma
  model,''
  Int.\ J.\ Mod.\ Phys.\  A {\bf 12} (1997) 1551
  [arXiv:hep-th/9606025].

\bibitem{Albertsson:2001dv}
  C.~Albertsson, U.~Lindstrom and M.~Zabzine,
  ``N = 1 supersymmetric sigma model with boundaries. I,''
  Commun.\ Math.\ Phys.\  {\bf 233} (2003) 403
  [arXiv:hep-th/0111161].

\bibitem{Yano1}
K.~Yano, ``Differential geometry on complex and almost 
 complex spaces'' (Pergamon, Oxford, 1965)

\bibitem{Yano2}
K.~Yano and M.~Kon, ``Structures of manifolds,''
 Series in Pure Mathematics, Vol.3
(World Scientific, Singapore, 1984)

\bibitem{Hull:1993kf}
  C.~M.~Hull,
  ``Lectures on W gravity, W geometry and W strings,''
  arXiv:hep-th/9302110.

\bibitem{Stojevic:2006pq}
  V.~Stojevic,
  ``Special holonomy and two dimensional supersymmetric sigma-models,''
  arXiv:hep-th/0611255.

\bibitem{Belov:2007qj}
  D.~M.~Belov, C.~M.~Hull and R.~Minasian,
  ``T-duality, Gerbes and Loop Spaces,''
  arXiv:0710.5151 [hep-th].

\bibitem{Strominger:1996it}
  A.~Strominger, S.~T.~Yau and E.~Zaslow,
  ``Mirror symmetry is T-duality,''
  Nucl.\ Phys.\  B {\bf 479} (1996) 243
  [arXiv:hep-th/9606040].
  
\bibitem{Stojevic:2008hc}
  V.~Stojevic,
  ``Doubled Formalism, Complexification and Topological Sigma-Models,''
  arXiv:0809.4034 [hep-th].




\bibitem{Grana:2008yw}
  M.~Grana, R.~Minasian, M.~Petrini and D.~Waldram,
  ``T-duality, Generalized Geometry and Non-Geometric Backgrounds,''
  JHEP {\bf 0904} (2009) 075
  [arXiv:0807.4527 [hep-th]].






  
\bibitem{Halpern:1995js}
  M.~B.~Halpern, E.~Kiritsis, N.~A.~Obers and K.~Clubok,
  ``Irrational conformal field theory,''
  Phys.\ Rept.\  {\bf 265} (1996) 1
  [arXiv:hep-th/9501144].
  
\bibitem{Halpern:1990cc}
  M.~B.~Halpern and N.~A.~Obers,
  ``Unitary Irrational Central Charge on Compact G: High Level Analysis and
  SU(3) Basic,''
  Nucl.\ Phys.\  B {\bf 345} (1990) 607.

  
  
\bibitem{Shelton:2006fd}
  J.~Shelton, W.~Taylor and B.~Wecht,
  ``Generalized flux vacua,''
  JHEP {\bf 0702} (2007) 095
  [arXiv:hep-th/0607015].

\end{thebibliography}
\end{document}